\documentclass[12pt]{article}
\usepackage{amssymb,graphicx} 
\usepackage{amsmath}
\usepackage{epsfig}

\def\e{{\rm e}}

\def\d{\partial}
\def\l{\left(}
\def\r{\right)}

\newcommand{\be}{\begin{equation}}
\newcommand{\ee}{\end{equation}}
\newcommand{\bea}{\begin{eqnarray}}
\newcommand{\eea}{\end{eqnarray}}
\newcommand{\bg}{\begin{gather}}
\newcommand{\eg}{\end{gather}}
\newcommand{\bseq}{\begin{subequations}}
\newcommand{\eseq}{\end{subequations}}

\renewcommand{\ln}{\mathop{\rm ln}\nolimits}
\newcommand{\sm}[1]{{\scriptscriptstyle \rm #1}}
\newcommand{\Tr}{{\rm Tr}}

\newcommand{\eq}[1]{(\ref{#1})}

\newcommand{\bra}[1]{\langle #1 |}
\newcommand{\ket}[1]{| #1 \rangle}

\begin{document}
\begin{center}
  {\Large\bf Domain walls between gauge theories} \\
\medskip  
S.L.~Dubovsky, S.M.~Sibiryakov\\
\medskip
  {\small
Institute for Nuclear Research of
         the Russian Academy of Sciences,\\  60th October Anniversary
  Prospect, 7a, 117312 Moscow, Russia
  }

\end{center}

\begin{abstract}
Noncommutative $U({\cal N})$ 
gauge theories at different ${\cal N}$ may be often thought of as different
sectors of a single theory: the $U(1)$ theory possesses a
sequence of vacua labeled by an integer parameter ${\cal N}$, and 
the theory in the vicinity of the ${\cal N}$-th vacuum coincides
with the $U({\cal N})$ noncommutative gauge theory. 
We construct noncommutative domain walls on fuzzy cylinder,
separating vacua with different gauge
theories. These domain
walls are solutions of BPS equations in gauge theory with an extra
term stabilizing the radius of the cylinder. We study properties of
the domain walls using adjoint scalar and fundamental fermion fields
as probes. We show that the regions on different sides of the wall
are not disjoint even in the low energy regime --- there are modes 
penetrating from one region to the other. We find that the wall
supports a chiral fermion zero mode. Also, we study non-BPS solution
representing a wall and an antiwall, and show that this solution is
unstable. We suggest that the domain
walls emerge as solutions of matrix model in large class of pp-wave
backgrounds with inhomogeneous field strength. In the M-theory language,
the domain walls have an interpretation of a stack of branes of
fingerstall shape inserted into a stack of cylindrical branes.
\end{abstract}

\section{Introduction}

Recently, field theories on noncommutative (NC) spaces attracted 
considerable interest (see, e.g.,
Refs.~\cite{konechnyshwartz,harvey,Douglas:2001ba,arefeva} for recent
reviews and references to earlier works).  One of the reasons for this
interest is the fact that NC gauge theories are non-local and
that the group of gauge transformations in NC theories contains some
of the diffeomorphisms\footnote{Strictly speaking, the latter property
holds only for fields in the adjoint representation of gauge
group.}. Both non-locality and invariance under diffeomorphisms are
expected to hold in the theory of quantum gravity, so one may hope to
gain some insight into techniques appropriate in quantum gravity,
using NC theories as a toy model (recall that NC theories are simpler in
the sense that non-locality is present there already at the classical
level).  This hope is further supported by the fact that NC gauge
theories emerge as effective descriptions of string theory in a
certain limit \cite{Seiberg:1999vs}, and are inherent in the matrix
approach to M-theory.

Among the consequences of non-locality and invariance under coordinate
transformations is the background dependence of both the matter content and
space-time interpretation of NC gauge theory.  One illustration of
this background dependence is that in many cases NC gauge theories
with different gauge groups $U({\cal N})$ emerge as different sectors
of a single theory
\cite{Gross:2000ss,Bak:2000zm,Bak:2001kq,Demidov:2003xq}.  For
instance, it was pointed out in Ref.~\cite{Gross:2000ss} that $U(1)$
gauge theory on the NC plane possesses a sequence of vacua labeled by a
natural number ${\cal N}$ with the following peculiar properties:
\newline 
i) Every vacuum with ${\cal N}>1$ is a highly
non-local field configuration from the point of view of the trivial
(${\cal N}=1$) vacuum;
\newline 
ii) Perturbation theory in the
vicinity of the ${\cal N}$-th vacuum is equivalent to perturbation
theory of the $U({\cal N})$ NC gauge theory above its
trivial vacuum;
\newline 
iii) The fact that there are different gauge
theories in different vacua cannot be understood as Higgs
mechanism. Namely, the action in
the vicinity of the ${\cal N}$-th vacuum contains $U({\cal N})$ gauge
fields only, with no extra massive vector bosons or Higgs fields.

The precise physical meaning of this phenomenon is not clear
yet. In the case of NC plane, it was argued \cite{Gross:2000ss} that 
${\cal N}$ is a superselection parameter, implying that
different sectors are completely disconnected from each other.

One way to understand the physical meaning of the 
parameter ${\cal N}$ is to study
whether there exist domain walls separating vacua with different
values of ${\cal N}$.
These walls, if any, are expected to exhibit interesting
physical properties, since in the low-energy (commutative) limit
they would serve as boundaries 
between regions with different gauge theories inside. On the other
hand, the absence of the domain walls would imply that 
sectors with different ${\cal N}$ are disconnected, and ${\cal N}$ is indeed
just a superselection parameter.
%

In this paper we study the problem of existence of the domain walls in
the context of gauge theory on a fuzzy cylinder introduced in
Ref.~\cite{Bak:2001kq} (see also Ref. \cite{Chaichian:2000ia}). 
We choose cylindrical geometry because in this case a domain wall
normal to the axis of the cylinder has  finite energy.
Fuzzy cylinder can be thought of as a lattice version of the
continuous NC cylinder studied in Ref.~\cite{Demidov:2003xq}. 
It is a well-defined NC space
which reduces to ordinary cylinder in the commutative limit.
An advantadge of the fuzzy cylinder setup 
compared to the continuous NC cylinder is that
the intriguing phenomenon described above
(existence of the sequence of vacua corresponding to different gauge theories) 
is more transparent in the former case. 
We elaborate on the relation
between the fuzzy and NC cylinders in section \ref{2}. 

We find that though configurations of finite energy interpolating
between different gauge vacua exist already in pure Yang--Mills
theory on the fuzzy cylinder (these configurations were first
constructed in Ref.~\cite{Bak:2001gm} and were interpreted there as
D-brane junctions), they are unstable and tend to dissolve classically
\footnote{In
Ref.~\cite{Bak:2001gm} it was suggested that these configurations may be
stabilized by quantum effects.}. However, once a gauge-invariant term,
stabilizing the radius of the cylinder, is added to the action, the
domain walls become stable solutions. They  saturate the BPS bound
emerging due 
to the existence of a non-trivial topological charge. 
We construct explicitly a family of solutions of the
 BPS equations, that describe domain walls between $U({\cal
N}_1)$ and $U({\cal N}_2)$ vacua with arbitrary ${\cal N}_1$ and
${\cal N}_2$. 

We suggest also a matrix model interpretation of these domain walls.
Namely, we 
argue that a term needed to stabilize the domain wall is generated in a
large class of curved gravitational pp-wave backgrounds with
inhomogeneous three-form field strength.
In the M-theory language, domain wall solutions have an interpretation
of a stack of branes of fingerstall shape, 
inserted into a stack of cylindrical branes.

An interesting property of the domain walls we construct is that the
regions with different 
gauge theories they separate are not disjoint even in the low energy (long
wavelength) regime. We illustrate this point 
by studying the properties of adjoint
scalar and fundamental 
fermion fields in the background of the simplest domain
wall between $U(1)$ and $U(2)$ gauge vacua. We show that there are
long wavelength modes of these fields, which penetrate from the
region with $U(2)$ gauge theory to that with $U(1)$ gauge theory. 
It is worth stressing that these modes do not belong to the diagonal
part of the $U(2)$ group.
Other modes, on the contrary, experience total reflection from the wall. In
addition, we find that the wall localizes a zero fermionic mode, whose
profile can be used as a probe of the shape of the wall.

Finally, we study a wall--antiwall
system 
which separates the cylinder into three regions with $U(1)$, $U(2)$ and
again $U(1)$ gauge theories. An unusual property of this
system (which is not BPS) is the absence of 
attraction between the wall and antiwall in the sense that there is a
one-parameter family of solutions with equal energies and different
distances between the wall and antiwall. However, a tachyonic
mode is present in the spectrum of excitations of the wall-antiwall
system. The absolute value of its 
mass squared is exponentially small when the distance between
the walls is large. Thus, the wall--antiwall system is almost stable,
when the walls are well separated.

This paper is organized as follows. In section \ref{2} we introduce
the algebra of functions on the fuzzy cylinder and describe the gauge
theory. In section \ref{3} we discuss the topology of domain
wall configuration, derive the BPS bound for its energy and solve the
corresponding BPS equations. In section \ref{4} we consider scalar and
fermion fields in the domain wall background. Wall--antiwall system is
studied in section \ref{wall-antiwall}. In section \ref{matrix} we
discuss the relation between our model and matrix theory in curved
background. The concluding section \ref{discussion} contains brief summary and
discussion of our results.

\section{Fuzzy cylinder}
\label{2}
\subsection{Algebra of functions on the fuzzy cylinder}
To introduce algebra ${\cal A}_C$ of functions on the NC
cylinder and algebra ${\cal A}_F$ of functions on the fuzzy cylinder,
it is convenient to start with the ``master'' algebra ${\cal A}$
generated by three
elements ${\bf x}$, ${\bf y}$, ${\bf z}$, obeying the following
commutation relations
\[
[{\bf z}, {\bf x}] = il{\bf y}~, ~~~[{\bf z}, {\bf y}] = -il{\bf x}~, ~~~  
[{\bf x}, {\bf y}] = 0 
\]
where the parameter $l$ is the scale of noncommutativity and
operators ${\bf x}$, ${\bf y}$ and ${\bf z}$ may be thought of as the
coordinates of a three-dimensional NC space where
NC and fuzzy cylinders are embedded into. 
It is convenient to introduce linear combinations 
\[
{\bf x_+} = {\bf x} + i{\bf y}~ , ~~~{\bf x_-} = {\bf x} - i{\bf y}\;.
\]
Commutation relations for these elements have the form 
\[
[{\bf z}, {\bf x_+}] = l{\bf x_+}~, ~~~[{\bf z}, {\bf x_-}] = -l{\bf
x_-}~, ~~~ 
[{\bf x_+}, {\bf x_-}] = 0 \;.
\]
Irreducible representations of the algebra ${\cal A}$ are labeled by the
eigenvalues of the two central elements
\[
T_1={\bf x_+}{\bf x_-}={\bf x}^2+{\bf y}^2~,~~~T_2 =
\e^{2\pi i{\bf{z}}/l}\;.
\]   
The eigenvalue of the first central element $T_1$ is natural to interprete as
radius squared of a cylinder. Thus, to introduce algebra 
${\cal A}_C$ of functions on the NC cylinder of radius
$\rho$, one sets
\[
T_1=\rho^2\;.
\]
More formally, algebra ${\cal A}_C$ can be defined as a factor-algebra
\[
{\cal A}_C={\cal A}/\{ T_1-\rho^2\}\;,
\]
where $\{ T_1-\rho^2\}$ is a subalgebra of algebra ${\cal A}$
generated by the element $( T_1-\rho^2)$. For recent study of gauge
theory on the NC cylinder see Ref.~\cite{Demidov:2003xq}.

On the other hand, by fixing the value of the second central element
\[
T_2=\e^{2\pi iz_0/l}
\]
one obtains a collection of planes parallel to $(x,y)$-plane with
$z$-coordinate equal to $(z_0+nl)$, $n\in \mathbb{Z}$. Algebra
${\cal A}_F$  of the fuzzy cylinder (see, e.g.,
Refs.~\cite{Bak:2001kq,Chaichian:2000ia}) is
obtained by fixing the values of both central elements $T_1$ and $T_2$. In
other words, one defines algebra ${\cal A}_F$ as the following
factor-algebra, 
\[
{\cal A}_F={\cal A_C}/\{ T_2-1\}={\cal A}/\{ T_1-\rho^2, T_2-1\}\;,
\]
where without loss of generality we set $z_0=0$. 
This algebra may be realized as 
the algebra of operators acting in a Hilbert space $H$ with
basis vectors $\ket{n}, n \in {\mathbb Z}$,
\be
{\bf z} = l\sum_{n=-\infty}^{\infty} n\ket{n}\bra{n}~,~~~
{\bf x_+} = \rho\sum_{n=-\infty}^{\infty} \ket{n+1}\bra{n}~,~~~
{\bf x_-} = \rho\sum_{n=-\infty}^{\infty} \ket{n-1}\bra{n}\;.
\label{cyl}
\ee
It is clear from the above discussion that fuzzy cylinder can be
thought of as a cylindrical semi-lattice with continuos coordinate
$\theta$ defined by
\[
x_\pm=\rho\e^{\pm i\theta}
\]
and discrete coordinate $z$ with spacing $l$.

To make this picture more transparent it is instructive to introduce
symbols of the operators in the Hilbert space $H$.  We will use the
symmetric ordering which maps functional exponents into operator
exponents
\[
\e^{i\l kz+N\theta\r}\to \e^{i\l k{\bf{z}}+N{\bf
\boldsymbol{\theta}}\r}=\e^{-ikNl/2}\e^{ik{\bf z}}\left(\frac{{\bf x}_+}{\rho}\right)^N
\;.
\]
Then it is straightforward to obtain the following relation between
an arbitrary operator 
$f=\sum f_{nm}\ket{n}\bra{m}$ and its symbol $\tilde{f}(z,\theta)$
\be
\tilde{f}(z,\theta)=l\sum_{n,m=-\infty}^{\infty}
\int_{-\frac{\pi}{l}}^{\frac{\pi}{l}} \frac{dk}{2\pi}~ f_{nm}~
\e^{ik\left(z-\frac{l(n+m)}{2}\right)+i(n-m)\theta}
\label{Weylsymb}
\ee 
Note that the integration over $k$ has finite range
$k\in (-{\pi}/{l}, {\pi}/{l})$. Consequently,
the symbol $\tilde{f}(z,\theta)$ is uniquely
determined by its values on the cylindrical
semi-lattice with lattice points
\[
z=nl\;,\;\;\;n\in \mathbb{Z}\;.
\]
Equivalently, one may view the map (\ref{Weylsymb}) as a
correspondence  between operators in $H$ and functions on the
cylinder, whose Fourier components along $z$-coordinate are cut off
at the scale $\pi/l$. Clearly, a symbol $\tilde{f}(z,\theta)$
considered as a function of the continuous variable $z$ contains the
same amount of information as its values on the lattice
$\tilde{f}(ln,\theta)$, but many formulae simplify when written in
terms of functions of the continuous variable.

The map \eq{Weylsymb} implies the following 
relation between trace of the operator 
and integral of the symbol,
\be
2\pi\rho l\Tr{f}=\int \rho d\theta dz\tilde{f} = \rho
l\sum_n\int d\theta\tilde{f}(ln,\theta)\;.
\label{integral}
\ee
Let us now define derivatives of functions on the fuzzy cylinder.
It is straightforward to check that differentiation of symbols with
respect to $\theta$-coordinate translates into the following inner
derivation in the operator language,
\be
\label{zderiv}
\partial_3\tilde{f}\equiv{i\over l}\widetilde{[{\bf
z}, f]}=\frac{\partial\tilde{f}}{\partial\theta}\;.
\ee
This relation has the same form as in the case of NC
cylinder (see, e.g. Ref. \cite{Demidov:2003xq}). However, it is impossible to define
a derivative along $z$-direction. Indeed, in the language  of symbols
we consider functions on the lattice in $z$ direction, so it is
natural to expect that some discretized version of the derivative in
$z$ direction emerges.
In the operator language, a naive attempt to define $z$-derivative
would contradict the Leibnitz rule because of the constraint $T_2=1$. 

On the other hand,
from the algebraic point of view it is natural to consider on 
equal footing the following three inner derivations of the algebra
 ${\cal A}_F$
\[
\partial_1 f=\frac{i}{l}[{\bf x},f]~,~~~
\partial_2 f=-\frac{i}{l}[{\bf y},f]~,~~~
\partial_3 f=\frac{i}{l}[{\bf z},f]\;.
\]
As pointed out above, the last derivation $\d_3$ corresponds
to differentiation with respect to $\theta$ in terms of symbols.
Furthemore, it is straightforward to check that
\bseq
\label{Weylderiv}
\begin{gather}
\partial_1\tilde{ f}\equiv\frac{i}{l}\widetilde{[{\bf x}
,f]}=y d_z\tilde{f}\label{Weylder1}\\
\partial_2 \tilde{f}\equiv-\frac{i}{l}\widetilde{[{\bf y},f]}
=x d_z\tilde{f}\;,
\label{Weylder2}
\end{gather}
\eseq
where finite-difference derivative $d_z$ is defined as 
\[
d_z\tilde{f}=\frac{\tilde{f}\left(z+{l}/{2}\right)
-\tilde{f}\left(z-{l}/{2}\right)}{l}\;.
\]
Note that this derivative has a simple form when written in
terms of symbol $\tilde{f}(z)$ considered as a function of continuous
variable $z$. However, written as a lattice derivative this
operator has the following highly non-local form
\[
d_z\tilde{f}(nl)=-{1\over \pi l}\sum_k(-1)^kk{\tilde{f}\l\l n+k\r
l\r\over k^2-1/4}\;.
\]
This property illustrates the fact that the formulation in terms of
functions $\tilde{f}(z)$ of continuous variable is often more
convenient than the lattice formulation.

\subsection{Scalar field on the fuzzy cylinder}
To get accustomed to physics of fuzzy cylinder, let us consider
a free scalar field theory on it. 
The action has the following form
\be
S= \int dt ~2\pi\rho l~ \Tr{
\frac{1}{2}\left( (\partial_0\phi)^2-
\frac{1}{\rho^2} ((\partial_1\phi)^2 +
(\partial_2\phi)^2 +
(\partial_3\phi)^2)-m^2\phi^2\right)}\;.
\label{scmatrix}
\ee  
The spectrum of this theory can be determined in two different
ways. First, one can rewrite the action (\ref{scmatrix})
in terms of the symbol of the operator $\phi$. Using
Eqs.~(\ref{integral}) -- (\ref{Weylderiv}) one has
\be
S=\int dt~\rho d\theta dz
\frac{1}{2}\left((\partial_0\tilde{\phi})^2-
\frac{1}{\rho^2}(\partial_{\theta}\tilde{\phi})^2
-(d_z\tilde{\phi})^2-m^2\tilde{\phi}^2\right)\;.
\label{scsymbol}
\ee
Solutions of the field equations following from the action
\eq{scsymbol} have
the form of waves
\[
\tilde{\phi}\propto \e^{-i\omega
t+ikz+iN\theta}\;,\;\;\;k\in\l-{\pi\over l},{\pi\over l}\r
\]
with the dispersion relation
\be
\omega^2=\frac{N^2}{\rho^2}+\left(\frac{2}{l}\sin\frac{kl}{2}\right)^2+m^2\;.
\label{disp}
\ee
In the long wavelength limit $k\ll 1/l$ we recover the usual
dispersion relation for scalar waves on the cylinder.

It is instructive to obtain the dispersion relation \eq{disp} in the 
operator approach. Variation of
the action \eq{scmatrix} yields the following field equation
\[
-\partial_0^2 \phi=\frac{1}{\rho^2l^2}
\left([{\bf z},[{\bf z},\phi]]+{1\over 2}[{\bf x_+},[{\bf x_-},\phi]]+
{1\over 2}[{\bf x_-},[{\bf x_+},\phi]]\right)+m^2\phi^2\;.
\]
In components $\phi_{nm}$ of the operator $\phi=\sum
\phi_{nm}\ket{n}\bra{m}$, this equation takes the form
\be
-\partial_0^2\phi_{nm}=\frac{1}{\rho^2}(n-m)^2\phi_{nm}+
\frac{1}{l^2}(2\phi_{nm}-\phi_{n+1,m+1}-\phi_{n-1,m-1})+m^2\phi_{nm}\;.
\label{sccomp}
\ee
The system of equations \eq{sccomp} decomposes into independent
recursion equations along diagonals of the matrix $\phi$. This
corresponds to the Kaluza-Klein decomposition over the compact variable
$\theta$. Let us consider 
equation \eq{sccomp} 
along the $N$th diagonal, $n-m=N$,  describing the $N$th
KK-mode,
\[
-\partial_0^2\phi_{n}=\frac{1}{\rho^2}N^2\phi_{n}+
\frac{1}{l^2}(2\phi_{n}-\phi_{n+1}-\phi_{n-1})+m^2\phi_{n}\;.
\]
where we have set $\phi_n\equiv\phi_{n,n+N}$.
We immediately find that the Ansatz $\phi_n=\phi_0\e^{-i\omega t+ikln}$ 
is consistent with this equation,
yielding the dispersion relation \eq{disp}.  

\subsection{Gauge theory on the fuzzy cylinder}
Let us now describe $U({\cal N})$ gauge theory on the fuzzy
cylinder. To this end we consider a field $\psi$ transforming
under the fundamental representation of $U({\cal N})$. 
This field belongs to the direct
sum of ${\cal N}$ copies of the algebra ${\cal A}_F$
\[
\psi \in{\mathbb C}^{\cal
N}\otimes {\cal A}_F\equiv\bigoplus_{i=1}^{{\cal N}} {\cal A}_F \;.
\] 
Gauge transformations are defined as follows,
\[
\psi \to U\psi\;,
\]
where $U$ is a unitary operator acting in ${\mathbb C}^{\cal
N}\otimes {\cal A}_F$ (equipped with scalar product
$(\psi_1,\psi_2)=\Tr(\psi_1^+\psi_2)$). 
Covariant connection is introduced
in the following way
\begin{gather}
\nabla(\partial_1)\psi = \frac{i}{l}({\bf X}\psi-\psi{\bf x})
\nonumber\\
\nabla(\partial_2)\psi = -\frac{i}{l}({\bf Y}\psi-\psi{\bf y})
\nonumber\\
\nabla(\partial_3)\psi = \frac{i}{l}({\bf Z}\psi-\psi{\bf z})
\nonumber \;,
\end{gather}
where covariant coordinates ${\bf X}, {\bf Y}, {\bf Z}$ 
are Hermitian ${\cal N} \times {\cal N}$ matrices
with entries in ${\cal A}_F$. They transform under the adjoint 
representation of the gauge
group, ${\bf X}\to U{\bf X}U^+$, etc. 
The covariant strength tensor is defined in the usual way,\footnote{The
second term in the r.h.s. of Eq.~\eq{fstrength} may appear somewhat
unusual. In fact, it is present in the conventional field theory as
well.  In commutative field theories this term ensures that
$F_{ij}$ is a tensor function (not a differential operator) when
vector fields $\d_i$ and $\d_j$ do not commute.}
\be
F_{ij}=[\nabla(\partial_i),\nabla(\partial_j)]-\nabla([\partial_i,\partial_j])
\;.
\label{fstrength}
\ee
Thus, we obtain
\bseq
\label{Fij}
\begin{gather}
F_{12}=\frac{1}{l^2}[{\bf X},{\bf Y}]\\
F_{13}=-\frac{1}{l^2}([{\bf X},{\bf Z}]+il{\bf Y})
\label{F13}\\
F_{23}=\frac{1}{l^2}([{\bf Y},{\bf Z}]-il{\bf X})
\label{F23}\\
F_{0j}=\frac{i}{l}(\partial_0{\bf X}_j-i[A_0,{\bf X}_j])~,~~~
{\bf X}_j = {\bf X}, {\bf Y}, {\bf Z}\;.
\end{gather}
\eseq
Now, it is straightforward to write down the Yang-Mills action,
\be
S=\frac{2\pi\rho l}{g^2}\Tr\left(-\frac{1}{\rho^2}F_{0i}^2+
\frac{1}{2\rho^4}F_{ij}^2\right)\;,
\label{gaugeaction}
\ee
where summation is assumed over indices $i,j = 1,2,3$. 
To figure out the commutative limit of the theory, it is convenient to
decompose the
covariant coordinates in the following way,
\bseq
\label{decomp}
\bea
&&{\bf X}={\bf x}-\frac{l}{2}({\bf y}A_z + A_z{\bf y})+\frac{l}{2}({\bf
x}A_{\rho}+A_{\rho}{\bf x})\label{xdecomp}\\
&&{\bf Y}={\bf y}+\frac{l}{2}({\bf x}A_z + A_z{\bf x})+\frac{l}{2}({\bf
y}A_{\rho}+A_{\rho}{\bf y})\label{ydecomp}\\
&&{\bf Z}={\bf z}-lA_{\theta}\;.\label{athet}
\eea
\eseq
Substituting expressions \eq{decomp} into Eqs.~\eq{Fij} 
and taking the limit $l\to 0$,
one obtains
\begin{gather}
F_{12}=-i\rho^2 D_zA_\rho \nonumber\\
F_{13}=i\rho(\sin{\theta}F_{\theta
z}-\cos{\theta}D_{\theta}A_{\rho})\nonumber\\
F_{23}=i\rho(\cos{\theta}F_{\theta
z}+\sin{\theta}D_{\theta}A_{\rho})\nonumber\\ 
F_{01}=i\rho(-\sin{\theta}F_{tz}+\cos{\theta}D_tA_{\rho})\nonumber\\
F_{02}=i\rho(\cos{\theta}F_{tz}+\sin{\theta}D_tA_{\rho})\nonumber\\
F_{03}=-iF_{t\theta}\nonumber\;,
\end{gather}
where
\[
F_{\alpha\beta}=\partial_{\alpha}A_{\beta}-\partial_{\beta}A_{\alpha}
-i[A_{\alpha},A_{\beta}]~,~~~
D_{\alpha}A_{\rho}=\partial_{\alpha}A_{\rho}-i[A_{\alpha},A_{\rho}]~,~~~
\alpha,\beta = t,\theta,z\;.
\]
Thus, in the commutative limit one obtains the usual Yang-Mills theory on
the cylinder, coupled to an adjoint scalar field $A_{\rho}$. The 
presence of
such a scalar is a consequence of the fact that the algebra of the
fuzzy cylinder ${\cal A}_F$
possesses three independent derivations instead of two, and hence
there are three gauge fields in the NC gauge theory.

Now, let us note the following peculiar 
property of NC
gauge theory on the fuzzy cylinder\footnote{This property is generic for gauge
theories on a non-compact NC manifold.
In the case of NC plane it was discussed in
Ref.~\cite{Gross:2000ss}, for NC cylinder in Ref.~\cite{Demidov:2003xq};
for earlier discussions of this property for fuzzy cylinder see Refs.
\cite{Bak:2001kq,Bak:2001gm}.}. 
Consider $U(1)$ gauge theory. Then
for {\it any} natural ${\cal N}$, there exists a vacuum in this theory, such
that the theory above this vacuum is identical to $U({\cal N})$ gauge theory
above its trivial vacuum. 
Namely, the vacuum corresponding to $U({\cal N})$ theory 
has the form
\bseq
\label{ZX^N}
\begin{align}
&{\bf Z}=l\sum_{n=-\infty}^{\infty}n\sum_{a=1}^{\cal N}
\ket{n{\cal N}+a}\bra{n{\cal N}+a}\label{Z^N}\\
&{\bf X}_+=\rho\sum_{n=-\infty}^{\infty}\sum_{a=1}^{\cal N}
\ket{(n+1){\cal N}+a}\bra{n{\cal N}+a}\label{X^N}\\
&A_0=0\;.
\end{align}
\eseq   
It is straightforward to check, that the energy of configuration \eq{ZX^N}
is equal to zero, and thus it indeed describes a vacuum in the $U(1)$
theory. Equivalence between this vacuum and the trivial
vacuum in $U({\cal N})$ theory can be established using an isomorphism
$S:{\mathbb C}^{\cal N}\otimes
H\to H$, defined on the basis vectors as
\[
\ket{a}\otimes\ket{n}\stackrel{S}{\mapsto}\ket{n{\cal N}+a}
\]
which maps operators \eq{Z^N}, \eq{X^N} into the following
operators acting in a direct sum
of ${\cal N}$ Hilbert spaces
\bseq
\label{ZX^NN}
\begin{gather}
{\bf Z}=
l\l\sum_{n=-\infty}^{\infty}n
\ket{n}\bra{n}\r\cdot{\bf{1}}\label{Z^NN}\\
{\bf X}_+=\rho\l\sum_{n=-\infty}^{\infty}
\ket{(n+1)}\bra{n}\r\cdot{\bf{1}}\label{X^NN}\;,
\end{gather}
\eseq   
where ${\bf 1}$ stands for the unit ${\cal N}\times{\cal N}$ matrix.
Field configuration given by Eqs. \eq{ZX^NN} describes a trivial vacuum
in the $U({\cal N})$ theory. Using the same trick for the field
configurations describing fluctuations above the vacuum given by
Eqs. \eq{ZX^N}, one observes that the action governing these
fluctuations is equivalent to the action describing fluctuations of
the $U({\cal N})$ theory in the vicinity of the trivial vacuum.

Thus, the $U(1)$ gauge theory on the fuzzy cylinder has
an infinite set of vacua labeled by ${\cal N}=1,2,\dots$ It is
natural to wonder whether these vacua correspond to different
superselection sectors, or are different phases of one and the
same theory. In particular, one may ask whether there exist solutions
(domain walls)
interpolating between different vacua. In the next section we answer
affirmatively to this question and explicitly find such domain wall 
solutions. 

\section{Domain wall}
\label{3}
In what follows it is convenient to use 
the dimensionless
variables
\be
Z=\frac{1}{l}{\bf Z}~,~~~X_{\pm}=\frac{1}{\rho}{\bf X}_{\pm}\;.
\label{redef}
\ee
We search for a static solution of $U(1)$ gauge theory
on the fuzzy cylinder in the form of a domain
wall, separating vacua corresponding to different gauge
theories. We will work in the gauge $A_0=0$. 
For concreteness we first concentrate on the case of a
domain wall between $U(1)$ gauge theory in the region $z<0$ and $U(2)$
gauge theory in the region $z>0$. Then the asymptotics of the domain
wall are given by Eqs. \eq{cyl} and \eq{ZX^N}, respectively,
\bseq
\label{asympt}
\begin{gather}
Z=\sum n\ket{n}\bra{n}~,~~~
X_+=\sum\ket{n+1}\bra{n}~,~~~n\to -\infty\label{asympt-}\\
Z= \sum n(\ket{2n-1}\bra{2n-1}+\ket{2n}\bra{2n})~,~~~
X_+=\sum\ket{n+2}\bra{n}~,~~~n\to +\infty\label{asympt+}\;.
\end{gather}
\eseq
Let us note that field configurations of finite energy, possessing
asymptotics \eq{asympt} do exist \cite{Bak:2001gm}. A simple example is
\bseq
\label{conf}
\begin{align}
&Z=\sum_{n=-\infty}^{0}n\ket{n}\bra{n}+
\sum_{n=1}^{\infty}n(\ket{2n-1}\bra{2n-1}+\ket{2n}\bra{2n})\label{confZ}\\
&X_+=\sum_{n=-\infty}^{0}\ket{n+1}\bra{n}+\sum_{n=1}^{\infty}\ket{n+2}\bra{n}
\label{confX+}
\end{align}
\eseq
This configuration is not a static solution of the field 
equations for gauge theory on the fuzzy cylinder with 
action given by Eq.~\eq{gaugeaction}. Nevertheless, it demonstrates
non-trivial topological properties one may expect for the domain wall
solution. In NC field theory 
the role of topological invariants is played
by traces
of commutators. Indeed, the latter do not change under 
small variations of operators they depend on. For configuration \eq{conf} we
have
\be
\Tr[X_+,X_-]=-1\;.
\label{1top}
\ee
The existence of such topological charge suggests a strategy of the
search for the domain wall solution: one may try
to obtain a BPS bound for the energy
functional following from Eq.~\eq{gaugeaction} in the topological sector
determined by Eq.~\eq{1top}. However, as we will see below, the action
\eq{gaugeaction} as it stands does not admit BPS solutions with desired
properties --- the domain wall tends to dissolve. 

To stabilize
the domain wall, 
let us introduce the following additional term into the gauge theory
action
\be
S_m=\int dt~\frac{2\pi\rho
l}{g^2}\Tr\left(-\frac{m^2}{16l^2}(X_+X_-+X_-X_+-2)^2\right)\;.
\label{addaction} 
\ee
This term has simple physical
effect: it stabilizes the radius of the cylinder. 
In the commutative limit it becomes a mass term for the
adjoint scalar field $A_{\rho}$,
\[
S_m\to\int dt\rho d\theta dz
\left(-\frac{m^2}{g^2}A_{\rho}^2\right)\;\;\; {\mbox{as}}\;l\to 0\;. 
\]
With this term added, the static energy takes the following form,
\be
{\cal E}=\frac{2\pi}{g^2\rho l}\Tr\left(|[Z,X_+]-X_+|^2
+\frac{\lambda^2}{4}[X_+,X_-]^2+
\frac{\mu^2}{4}(X_+X_-+X_-X_+-2)^2\right)
\label{energyfunc}
\ee
where we introduced dimensionless parameters
\be
\lambda=\frac{\rho}{l}~,~~~\mu=\frac{m\rho}{2}\;.
\ee
This expression can be rewritten in the BPS form
\begin{multline}
{\cal E}=\frac{2\pi}{g^2\rho l}\Tr\left(|[Z,X_+]-X_+|^2+
\left(\frac{\lambda}{2}[X_+,X_-]\pm\frac{\mu}{2}(\{X_+,X_-\}-2)\right)^2
\right.\\
\left.\mp\frac{\lambda\mu}{4}\{[X_+,X_-],(\{X_+,X_-\}-2)\}\right)\ge
\mp\frac{2\pi}{g^2\rho l}\frac{\lambda\mu}{2}Q\;,
\label{BPSenergy}
\end{multline}
where braces stand for anti-commutators, and
\bea
Q=\frac{1}{2}\Tr\{[X_+,X_-],(\{X_+,X_-\}-2)\}
=\Tr[X_+,(X_-X_+X_--2X_-)]
\label{topcharge}
\eea
is a topological charge. For the configuration \eq{conf} the
topological charge is equal to one.
Energy in each topological sector is
minimized by the solution of BPS equations, which follow from
Eq. \eq{BPSenergy},
\bseq
\label{BPSeqns}
\begin{align}
&[Z,X_+]-X_+=0\label{BPSeqn1}\\
&[Z,X_-]+X_-=0\label{BPSeqn2}\\
&(\lambda+\mu)X_-X_+\pm(\lambda-\mu)X_+X_-=2\mu\label{BPSeqn3}\;.
\end{align}
\eseq
These equations can be considered as defining an algebra with three
generators $Z$, $X_+$, $X_-$. Each solution of Eqs. \eq{BPSeqns}
decomposes 
into direct sum of operators acting in
irreducible representations of this algebra. Thus, we have to classify
irreducible representations of the algebra \eq{BPSeqns}.

Let us work in the eigenbasis $\{|z\rangle\}$ of the operator $Z$,
\[
Z\ket{z}=z\ket{z}
\]
It follows from Eqs.~\eq{BPSeqn1}, \eq{BPSeqn2}, that $X_+$, $X_-$ raise
and lower the eigenvalue $z$ by one,
\[
X_+\ket{z}=x_+(z)\ket{z+1}~,~~~ X_-\ket{z}=x_-(z)\ket{z-1}\;.
\]
Hence, in an irreducible representation, the eigenvectors of $Z$ 
may be labeled by an integer $k$, and in this
basis one has
\begin{gather}
Z=\sum_k (k+\gamma)\ket{k}\bra{k}\nonumber\\
X_+=\sum_k x_k~\ket{k+1}\bra{k}\nonumber\\
X_-=\sum_k x^*_k~\ket{k}\bra{k+1}\nonumber\;,
\end{gather}
where $\gamma$ is a real parameter characteristic to the irreducible
representation.
Relation \eq{BPSeqn3} gives a recursion equation for the coefficients
$x_k$. Let us consider the lower sign in Eq.~\eq{BPSeqn3}, then
\[
(\lambda+\mu)|x_k|^2-(\lambda-\mu)|x_{k-1}|^2=2\mu\;.
\]
The solution of this equation is
\[
|x_k|^2=1+C\alpha^k
\]
where
\[
\alpha=\frac{\lambda-\mu}{\lambda+\mu}\;,
\]
and $C$ is an arbitrary constant. 
Taking into account that $|\alpha|<1$ and that  $|x_k|^2>0$ 
we obtain three  different possibilities (up to redefinitions of $k$,
$C$ and $\alpha$): \\
{\em a}) $C=0$ and $k$ runs from $-\infty$ to $+\infty$; 
in this case $x_k=1$,\\  
{\em b}) $C=-1$ and $k$ runs from 1 to $+\infty$; in this case 
$x_k=\sqrt{1-\alpha^k}$,\\
{\em c}) $C>0$ and $\alpha>0$ and $k$ runs from $-\infty$ to $+\infty$.

One may give the following geometric interpretation to these three types
of solutions. Let us consider them as axially symmetric fuzzy
manifolds where the role of coordinates is played by covariant
coordinates $X_+$, $X_-$ and $Z$. This kind of interpretation is quite common
in NC gauge theories; later it will be supported by the
study of the spectrum of small perturbations. The case  ({\em a})
corresponds just to the original fuzzy cylinder (see Fig. 1a)). The
case ({\em b})
corresponds to the semi-infinite fuzzy cylinder (``fuzzy fingerstall'') 
of variable radius,
starting at $z=\gamma$ and oriented towards positive values of
$z$ (see Fig. 1b)). Finally, the case ({\em c}) corresponds to an infinite
tube whose radius tends to unity as $z\to +\infty$ and becomes infinite
as $z\to -\infty$ (see Fig. 1c)). 
\begin{figure}[htb]
\begin{center}
\begin{picture}(500,100)(0,0)
\put(10,20){\epsfig{file=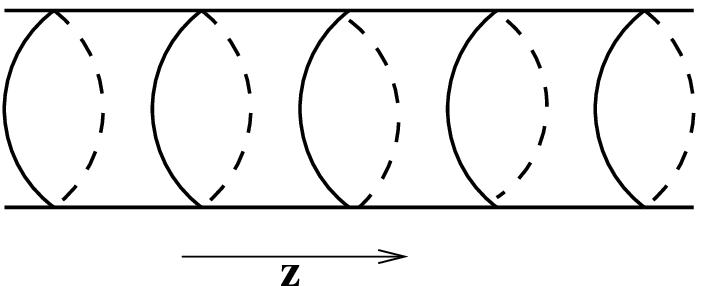,height=2.cm,width=4.cm} }
\put(170,20){\epsfig{file=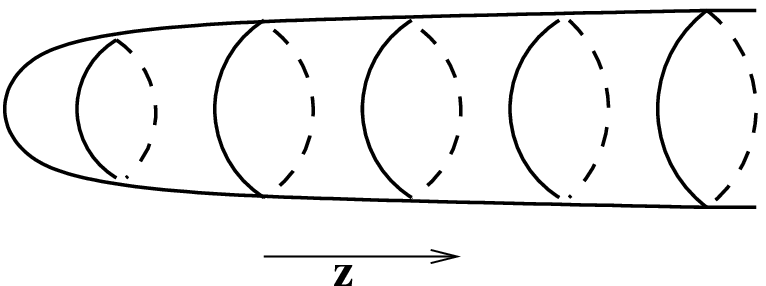,height=2.cm,width=4.cm}}
\put(340,10){\epsfig{file=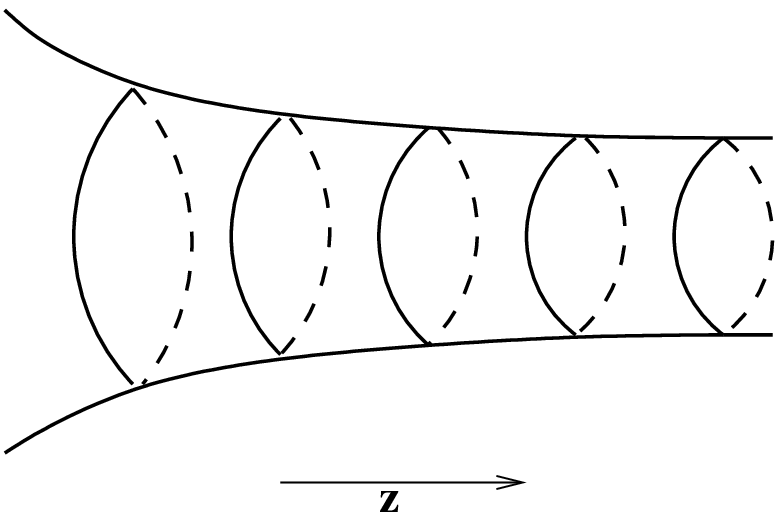,height=3.cm,width=4.cm}}
\put(55,0){a}
\put(218,0){b}
\put(394,0){c}
\end{picture}
\end{center}
\caption{
Geometrical interpretation of the three types of solutions to
BPS equations \eq{BPSeqns} \newline
a) fuzzy cylinder, b) fuzzy fingerstall, c)
infinitely expanding tube.}
\end{figure}
It is straightforward to check that the topological charge $Q$ is equal to
zero in the first case, equal to one in the case ({\em b}) and is 
ill-defined in the case ({\em c}). We will not consider solutions of the
third type in this paper.

The solution corresponding to a domain wall between $U(1)$ and $U(2)$
gauge theories is a direct sum of representations ({\em a}) and ({\em b}),
\bseq
\label{domwall}
\begin{align}
&Z=\sum_{n=-\infty}^{0}n\ket{n}\bra{n}+
\sum_{n=1}^{\infty}n\ket{2n-1}\bra{2n-1}+
\sum_{n=1}^{\infty}(n+\gamma)\ket{2n}\bra{2n}\label{domwall1}\\
&X_+=\sum_{n=-\infty}^{0}\ket{n+1}\bra{n}+
\sum_{n=1}^{\infty}\ket{2n+1}\bra{2n-1}+
\sum_{n=1}^{\infty}\sqrt{1-\alpha^n}\ket{2n+2}\bra{2n}\label{domwall2}\;.
\end{align}
\eseq
This domain wall has a simple geometrical  interpretation: 
it describes a parallel system of an 
infinite fuzzy cylinder and a half-infinite fuzzy fingerstall (see Fig. 2).
The topological charge $Q$ of this solution is equal to unity, as expected.
Without loss of generality we dropped in Eq.~\eq{domwall1} the possibility
of the overall shift of $Z$ by a $c$-number. Parameter
$\gamma$, as we will show later,
characterizes the position of the wall along $z$-coordinate. 
The energy of the wall is
\be
{\cal E}=\frac{\pi m\rho}{2g^2l^2}
\label{dmwenergy}
\ee
Its width $\delta z$ can be estimated as the size of the region along
$z$-direction, where
exact solution \eq{domwall} differs considerably from  
its asymptotic form \eq{asympt}.
Then, restoring the dimensionality, one has
\be
\delta z =\frac{l}{|\ln{\alpha}|} \approx \frac{1}{m}\;\;\;{\mbox{at}}\;
m\ll\frac{1}{l}\;.
\label{dmwwidth}
\ee
Note, that as $m$ tends to zero, the width of the wall becomes
infinite. This justifies our previous claim that in NC pure
Yang--Mills theory, the wall
tends to dissolve. Thus, we keep $m\neq 0$.
In the commutative limit $l\to 0$ the energy of the wall diverges,
 while its width stays
finite. Another interesting limit is that of the NC
plane. It corresponds to $\rho\to\infty$, $l\to 0$, and
$\vartheta=\rho l$ fixed. In this regime the energy \eq{dmwenergy}
diverges as $\rho^3$, which means that our solution does not
correspond to a domain wall of finite tension on the NC
plane.
\begin{figure}[tb]
\begin{center}
\epsfig{file=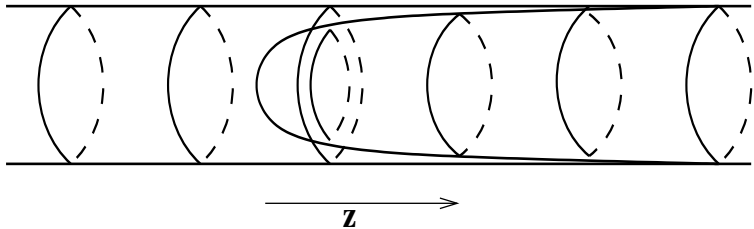,height=3.cm,width=10.cm} 
\end{center}
\caption{Geometrical interpretation of the domain wall: fingerstall
inserted into the fuzzy cylinder.}
\end{figure}

The solution \eq{domwall} is easy to generalize to $(U({\cal N}_1)-U({\cal
N}_2))$ domain wall with ${\cal N}_1 < {\cal N}_2$. One simply takes
the direct sum of ${\cal N}_1$ representations of type ({\em a})
(cylinders) and
${\cal N}_2-{\cal N}_1$ representations of type ({\em b})
(fingerstalls). 
The antiwall (that is $(U({\cal N}_1) - U({\cal N}_2))$ wall with 
${\cal N}_1 > {\cal N}_2 $) is the direct sum of irreducible
representations of the algebra defined by relations \eq{BPSeqns} with
upper sign in \eq{BPSeqn3}. A system of a wall and an
antiwall will be studied in detail in section \ref{wall-antiwall}. 

\section{Scalars and fermions in the domain wall background}
\label{4}
\subsection{Adjoint scalar}
To figure out the properties of the domain wall solution constructed
in the previous section, let us consider a 
Hermitian adjoint scalar field $\phi$ in the $(U(1)-U(2))$ domain wall
background. The action is
\[
S_{\phi}=2\pi\rho l\int dt\, \Tr\left({1\over 2}{(D_0\phi)^2}+
\frac{1}{2\rho^2 l^2}\l
[{\bf Z},\phi]^2+
[{\bf X},\phi]^2+[{\bf Y},\phi]^2\r-
{1\over 2}{m_{\phi}^2\phi^2}\right)\;.
\]
After redefinition \eq{redef} one obtains
\be
S_{\phi}=\frac{\pi l}{\rho}\int dt\,\Tr\left(\rho^2(D_0\phi)^2+
[Z,\phi]^2-\lambda^2|[X_+,\phi]|^2-\mu_{\phi}^2\phi^2\right)\;.
\label{scact}
\ee
where $\mu_{\phi}=m_{\phi}\rho$. Equations for normal modes following
from Eq.~\eq{scact} are
\be
\rho^2\omega^2\phi=[Z,[Z,\phi]]+{\lambda^2\over 2}[X_-,[X_+,\phi]]+
{\lambda^2\over 2}[X_+,[X_-,\phi]]+\mu_{\phi}^2\phi\;,
\label{sceq}
\ee
where 
operators $Z$, $X_+$
are given by \eq{domwall}. It is convenient to decompose the Hilbert space
$H$ into direct sum $H=H_1\oplus H_2$ of two Hilbert spaces with
bases
\[
\ket{c_n}=\begin{cases}\ket{n} &n\leq 0\\
                       \ket{2n-1} &n\geq 1
          \end{cases}
\] 
and
\[
\ket{h_p}=\ket{2p}~,~~~p\geq 1
\]
respectively. Background operators \eq{domwall} take the following form
\bseq
\label{backgrnd}
\begin{align}
&Z=\sum_{n=-\infty}^{\infty}n\ket{c_n}\bra{c_n}+
\sum_{p=1}^{\infty}(p+\gamma)\ket{h_p}\bra{h_p}\label{backgrnd1}\\
&X_+=\sum_{n=-\infty}^{\infty}\ket{c_{n+1}}\bra{c_n}+
\sum_{p=1}^{\infty}\sqrt{1-\alpha^p}\ket{h_{p+1}}\bra{h_p}\;.\label{backgrnd2}
\end{align}
\eseq
These operators are block-diagonal in the sense that they 
map $H_i$ to $H_i$ ($i=1,2$). It is convenient to decompose 
the field $\phi$ as
\be
\phi=\sum\phi_{nm}\ket{c_n}\bra{c_m}+\sum\varphi_{pq}\ket{h_p}\bra{h_q}+
\sum(\chi_{np}\ket{c_n}\bra{h_p}+\chi_{np}^*\ket{h_p}\bra{c_n})\;,
\label{phidecomp}
\ee
so that equations for different components $\phi_{nm}$, $\varphi_{pq}$
and $\chi_{np}$ decouple. 
One immediately notices 
that the indices $p$, $q$ are greater than zero, so nonvanishing
matrix elements of the operators $\varphi$ and $\chi$ have at least
one positive index. This implies that the modes described by these
operators live in the region $z > 0$ and cannot penetrate into the
region $z\to -\infty$.
Let us study the three types of modes entering the decomposition
\eq{phidecomp}
separately.  

Equations for the $\phi_{nm}$-components in the
background \eq{backgrnd} exactly coincide with equations
for the scalar field on the fuzzy cylinder \eq{sccomp}. 
Thus, excitations of this
type do not feel the presence of the fingerstall and freely propagate
from one end of the cylinder to the other.

Equations for the $\varphi_{pq}$-components have the following form,
\begin{align}
\rho^2\omega^2\varphi_{pq}&=
\left((p-q)^2+\mu_{\phi}^2+2\lambda^2
\left(1-\frac{\alpha^p+\alpha^{p-1}+\alpha^q+\alpha^{q-1}}{4}\right)
\right)\varphi_{pq}\nonumber\\
&-\lambda^2\sqrt{(1-\alpha^p)(1-\alpha^q)}\,\varphi_{p+1,q+1}
-\lambda^2\sqrt{(1-\alpha^{p-1})(1-\alpha^{q-1})}\,\varphi_{p-1,q-1}\;.
\nonumber
\end{align}
Let us perform Kaluza-Klein decomposition
in analogy to the case of free scalar field considered in section
\ref{2}. Namely, after fixing the number of the Kaluza--Klein mode,
$p-q=N$, we obtain the following
recursion relation for $\varphi_q\equiv\varphi_{q+N,q}$,
\begin{align}
\rho^2\omega^2\varphi_q&=
\left(N^2+\mu_{\phi}^2+2\lambda^2
\left(1-\frac{\alpha^{N+q}+\alpha^{N+q-1}+\alpha^q+\alpha^{q-1}}{4}\right)
\right)\varphi_q\nonumber\\
&-\lambda^2\sqrt{(1-\alpha^{N+q})(1-\alpha^q)}\,\varphi_{q+1}
-\lambda^2\sqrt{(1-\alpha^{N+q-1})(1-\alpha^{q-1})}\,\varphi_{q-1}\;.
\label{qeq}
\end{align}
At large $q$, the general solution of this equation is
\be
\varphi_q=A\e^{iklq}+B\e^{-iklq}\;,
\label{qasymp}
\ee
where the wave vector $k$ and frequency $\omega$ are related by
Eq.~\eq{disp}. To relate the coefficients $A$ and $B$ let us note that
equation \eq{qeq} leads to the conservation of ``current'',
\[
J_{q+1}-J_q=0
\]
with
\[
J_q=i\sqrt{(1-\alpha^{N+q})(1-\alpha^q)}
(\varphi_{q+1}^*\varphi_q-\varphi_q^*\varphi_{q+1})\;.
\]
Equation \eq{qeq} with $q=1$ implies that
\[
\varphi_2=c\varphi_1
\]
with real coefficient of proprtionality $c$. So
$J_1=0$ and,  as a consequence, we obtain that $J_q=0$ for all $q$. Then
the coefficients in Eq.~\eq{qasymp} have equal absolute values,
\[
|A|=|B|\;.
\] 
The asymptotic solution \eq{qasymp} describes two waves with equal amplitudes
propagating in the opposite directions. In other words, excitations 
$\varphi_{pq}$ experience total reflection from the domain wall.
Note that these modes do not feel the presence of the fuzzy cylinder 
and live entirely on the fingerstall. 

Modes $\chi_{np}$ obey the following equation,
\begin{multline}
\rho^2\omega^2\chi_{np}=\left((n-p-\gamma)^2+\mu_\phi^2+2\lambda^2
\left(1-\frac{\alpha^p+\alpha^{p-1}}{4}\right)\right)\chi_{np}\\-
\lambda^2\sqrt{1-\alpha^{p-1}}\,\chi_{n-1,p-1}-
\lambda^2\sqrt{1-\alpha^p}\,\chi_{n+1,p+1}\;.
\nonumber
\end{multline} 
By fixing $n-p=N$ and substituting $\chi_p\equiv\chi_{p+N,p}$ one obtains
\begin{multline}
\rho^2\omega^2\chi_p=\left((N-\gamma)^2+2\lambda^2
\left(1-\frac{\alpha^p+\alpha^{p-1}}{4}\right)\right)\chi_p\\-
\lambda^2\sqrt{1-\alpha^{p-1}}\,\chi_{p-1}-
\lambda^2\sqrt{1-\alpha^p}\,\chi_{p+1}\;.
\nonumber
\end{multline} 
The analysis similar to that for modes $\varphi$ demonstrates that solutions of
this equation are waves which experience total reflection from the 
domain wall. Their dispersion relation is
\be
\omega^2=\frac{(N-\gamma)^2}{\rho^2}+
\left(\frac{2}{l}\sin{\frac{kl}{2}}\right)^2+m_{\phi}^2\;.
\label{chidisp}
\ee  
Let us comment on this result. 
First, if the parameter $\gamma$
is not integer, spectrum of the off-diagonal modes $\chi_{np}$ differs
from that of diagonal modes $\phi_{nm}$ and $\varphi_{pq}$. As 
$z\to +\infty$, the $\chi$ and $(\phi,\varphi)$ modes 
correspond to off-diagonal and
diagonal components of the adjoint $U(2)$ field, respectively. Difference
in their dispersion relations means that in the case of non-integer
$\gamma$ the gauge group $U(2)$  is broken down to
$U(1)\times U(1)$ at $z\to +\infty$. 
This effect is nothing else than spontaneous breaking of
the gauge symmetry by a nontrivial Wilson line on the
cylinder. Indeed, from
\eq{athet} and \eq{domwall1} we see that 
\[
A_{\theta}=\bigl(\begin{smallmatrix} 0&0\\ 0&\gamma
\end{smallmatrix}\bigr)~,~~~z\to +\infty\;.
\]
The value of the Wilson line around the cylinder is equal to
$\e^{2\pi i\gamma}\neq 1$ if $\gamma$ is not integer. Thus,
$A_{\theta}$ cannot be removed by a gauge transformation in this case
and breaks the gauge symmetry. Conversely, if $\gamma$ is integer,
$A_{\theta}$ is a pure gauge as $z\to +\infty$, and the 
symmetry $U(2)$ is unbroken
in the $z\to +\infty$ asymptotics. From now on we consider
integer values of $\gamma$ only.

Let us see now that the parameter $\gamma$ is related to the position of the
domain wall.
Equation \eq{chidisp} implies that the genuine Kaluza-Klein number of 
a $\chi$ excitation which characterizes its energy is
$N_0=N-\gamma$. The expression for $N_0$th KK
excitation in the operator form is
\[
\chi^{(N_0)}=\sum_{p=1}^{\infty}\chi_{p}^{(N_0)}\ket{c_{\gamma+N_0+p}}
\bra{h_p}
\] 
The operator $\chi^{(N_0)}(\chi^{(N_0)})^{\dag}$ acts diagonally in $H_1$
and annihilates all vectors with 
\[
n<n_{min}\equiv N_0+\gamma+1\;.
\]
Hence, the $\chi$-wave is reflected from a point with $z$-coordinate 
$n_{min}$. 
For a given KK mode this coordinate depends
additively on $\gamma$, implying that this parameter is natural
to interpret as the
position of the wall\footnote{Strictly speaking, the above argument is not 
rigorous. The reason is that bilinear combinations of adjoint fields are
again adjoints and are not gauge invariant, while 
in NC theory gauge transformations of adjoints
include change of coordinates. Our conclusion that $\gamma$
characterizes the position of the domain wall will be
confirmed by the consideration of a fundamental field in the next
subsection. 
}.

Let us summarize our results on the spectrum of the adjoint field in
the presence of the domain wall. In the region $z\to +\infty$ there
are four (real) modes which constitute adjoint representation of
$U(2)$. Their dispersion relation is given by Eq.~\eq{disp} which reduces
to the standard dispersion relation on the cylinder in the long
wavelength regime. Three of these modes are reflected from the wall
while one freely propagates to the region
$z\to -\infty$. A notable fact is that the latter mode 
does not decouple from the other modes in the region $z\to +\infty$: in
terms of $U(2)$ group it has the form 
$\bigl(\begin{smallmatrix}
\phi&0\\ 0&0
\end{smallmatrix}\bigr)$, and couples to other modes due to gauge
interactions. These results can be straightforwardly
generalized to the case of the $(U({\cal N}_1)-U({\cal N}_2))$ domain
wall with ${\cal N}_1<{\cal N}_2$. In that case there are ${\cal N}_1^2$
modes in the adjoint of $U({\cal N}_1)$ group freely propagating along
the cylinder and $({\cal N}_2^2-{\cal N}_1^2)$ modes exhibiting
reflection from the domain wall.

The analysis of gauge field perturbations in the domain wall
background is less transparent because of mixing between
different components and the necessity to implement the Gauss'
constraint. We do not give the details here, and only states that the
qualitative picture described in the previous paragraph holds for the
gauge field perturbations as well. 
Namely, the $U({\cal N}_1)$ gauge fields propagate freely
along the entire cylinder; in the region $z\to +\infty$ they become a
part of $U({\cal N}_2)$ gauge multiplet; the rest of the $U({\cal
N}_2)$ gauge fields live only in the region on the right and are totally 
reflected from the wall.

\subsection{Fundamental fermion}

To study fermions in the background of the domain wall, we
have to  define the Dirac operator on the fuzzy cylinder. To this end let us
consider possible choices of the Dirac Hamiltonian on the commutative
cylinder. The simplest one is
\[
\tilde{D}=i\sigma_2\partial_z-\frac{i}{\rho}\sigma_3\partial_{\theta}
+\sigma_1m_f
\]
where $m_f$ is the fermion mass and $\sigma_i$ are Pauli matrices. 
This operator is not
convenient for our purposes, 
because natural derivative
operators on the fuzzy cylinder 
are $x\partial_z$ and $y\partial_z$ rather than $\partial_z$
(see Eqs.~\eq{Weylder1}, \eq{Weylder2})
\footnote{Nevertheless, one could proceed with the operator
$\tilde{D}$ and rewrite the term $i\sigma_2\partial_z\psi$ as
\[ 
i\frac{\sigma_2}{\rho^2}(x\cdot x\partial_z\psi+y\cdot y\partial_z\psi)
\]
Noncommutative counterpart of this expression would be
\[
\frac{\sigma_2}{\rho^2 l}({\bf x}[{\bf y},\psi]-{\bf y}[{\bf
x},\psi])=
\frac{\sigma_2}{\rho^2 l}({\bf x}\psi{\bf y}-{\bf y}\psi{\bf x})
\]
This operator leads to the second order recursion equations and
its spectrum suffers from doubling of fermion species, resembling that 
occurring
in the lattice field theory.}.
So, we will work with the following
unitary equivalent operator
\[
D=S^+\tilde{D}S\;,
\]
where
\be
S=
\begin{pmatrix}
e^{i\theta/2}&0\\
0&e^{-i\theta/2}
\end{pmatrix}\;.
\label{matrixS}
\ee
Explicitly, one has
\be
D=\frac{1}{\rho}\left(-i(\sigma_1y-\sigma_2x)\partial_z
-i\sigma_3\partial_{\theta}+\frac{1}{2}+
(\sigma_1x+\sigma_2y)m_f\right)\;.
\label{DiracCom}
\ee
Let us note that due to the form \eq{matrixS} of the 
unitary transformation $S$, periodic boundary conditions for the operator
$D$ correspond to anti-periodic boundary conditions for $\tilde{D}$ and
vice versa. This is not problematic, as one may consider both
periodic and anti-periodic spinors on the cylinder.

Noncommutative counterpart of the operator \eq{DiracCom} can be naturally
defined in terms of its action on a
spinor field $\psi$ as (cf. Eqs. \eq{zderiv}, \eq{Weylderiv})
\be
D\psi=\frac{1}{\rho l}(\sigma_1[{\bf x},\psi]+\sigma_2[{\bf y},\psi]
+\sigma_3[{\bf z},\psi])+\frac{1}{2\rho}\psi+
\frac{m_f}{2\rho}(\sigma_1\{{\bf x},\psi\}+\sigma_2\{{\bf y},\psi\})\;.
\label{DiracNC}
\ee
The operator ordering in the mass term in Eq.~\eq{DiracNC} is chosen somewhat
arbitrarily; a different choice would not alter the results of the
analysis below but would make calculations more cumbersome. The spectrum
of the operator \eq{DiracNC} is 
\be
\omega^2=\frac{(N+1/2)^2}{\rho^2}+
\left(\frac{2}{l}\sin{\frac{kl}{2}}\right)^2+
\left(m_f\cos{\frac{kl}{2}}\right)^2\;,
\label{dispDirac}
\ee
where $N$ is the Kaluza-Klein number and $k$ is the wave vector. 
Two comments on this dispersion relation are in order. 
First, the appearance of $(N+1/2)$ instead of $N$ in the
first term of Eq.~\eq{dispDirac} is not unexpected, 
since as discussed above we are effectively 
considering anti-periodic fermions. The dependence on
$k$ in the third (mass) term is more surprising. 
Interestingly, the same dependence appears in the
dispersion relation for the field $A_{\rho}$ defined by Eqs.~\eq{xdecomp},
\eq{ydecomp} when the term \eq{addaction} is added to the gauge action. 

The Dirac Hamiltonian for fermion in the fundamental representation of
the gauge 
group is obtained from Eq.~\eq{DiracNC} via substitution
\begin{gather}
[{\bf x},\psi]~ \longrightarrow ~{\bf X}\psi-\psi {\bf x}\nonumber\\
\{{\bf x},\psi\}~ \longrightarrow ~{\bf X}\psi+\psi {\bf x}\nonumber
\end{gather}
and similarly for other coordinates. In the domain wall background
\eq{domwall}, it is convenient to decompose the fermion field in the
following way (cf. Eq.~\eq{phidecomp})
\[
\psi=\sum_{n,m=-\infty}^{\infty}\psi_{nm}\ket{c_n}\bra{m}+
\sum_{p=1, n=-\infty}^{\infty}\eta_{pn}\ket{h_p}\bra{n}\;.
\]
Equations for the $\psi_{nm}$-components are identical to those
obtained in the
case of  fuzzy cylinder without the domain wall, and the corresponding
branch of
the spectrum is given by Eq. \eq{dispDirac}. Thus,
these modes do not feel the presence of the domain wall at all. For
spinors $\eta_{pn}$ we obtain the following equations,
\[
\omega\rho\begin{pmatrix}\eta^1_{pn}\\
                         \eta^2_{pn}\end{pmatrix}=
\begin{pmatrix}
(p-n+\gamma+\frac{1}{2})\eta^1_{pn}+
(\lambda+\frac{\mu_f}{2})\sqrt{1-\alpha^p}\eta^2_{p+1,n}
-(\lambda-\frac{\mu_f}{2})\eta^2_{p,n-1}\\
(\lambda+\frac{\mu_f}{2})\sqrt{1-\alpha^{p-1}}\eta^1_{p-1,n}
-(\lambda-\frac{\mu_f}{2})\eta^1_{p,n+1}-(p-n+\gamma-\frac{1}{2})\eta^2_{pn}
\end{pmatrix},
\]
where $\mu_f=m_f\rho$. After
substituting $\eta^1_p\equiv \eta^1_{p,p+N}$,
$\eta^2_p\equiv \eta^2_{p,p+N-1}$,  equations along diagonals $n-p=N$ 
take the form
\[
\omega\rho\begin{pmatrix}\eta^1_{p}\\
                         \eta^2_{p}\end{pmatrix}=
\begin{pmatrix}
(-N+\gamma+\frac{1}{2})\eta^1_{p}+
(\lambda+\frac{\mu_f}{2})\sqrt{1-\alpha^p}\eta^2_{p+1}
-(\lambda-\frac{\mu_f}{2})\eta^2_{p}\\
(\lambda+\frac{\mu_f}{2})\sqrt{1-\alpha^{p-1}}\eta^1_{p-1}
-(\lambda-\frac{\mu_f}{2})\eta^1_{p}-(-N+\gamma+\frac{1}{2})\eta^2_{p}
\end{pmatrix}.
\]
Since $n=p+N > N$, a KK mode with fixed $N$ does not penetrate into 
the region $z\to
-\infty$ and thus is reflected from the domain wall. In this respect 
the situation
is similar to the case of the adjoint scalar
considered in the previous subsection. 

Let us find now a zero mode localized on the domain
wall. By zero mode we understand a mode which is
annihilated by the transverse part of the Dirac Hamiltonian.
In other words, it obeys the following equations
\begin{align}
\label{firsteta}
&\left(\lambda+\frac{\mu_f}{2}\right)\sqrt{1-\alpha^{p-1}}\,\eta^1_{p-1}-
\left(\lambda-\frac{\mu_f}{2}\right)\eta^1_{p}=0\\
\label{secondeta}
&\left(\lambda+\frac{\mu_f}{2}\right)\sqrt{1-\alpha^p}\,\eta^2_{p+1}-
\left(\lambda-\frac{\mu_f}{2}\right)\eta^2_{p}=0\;.
\end{align}
Equation \eq{firsteta} with $p=1$ implies that $\eta^1_1=0$, and as a
consequence
\be
\eta^1_p=0
\label{eta1}
\ee 
for all $p$.
From Eq.~\eq{secondeta} one obtains the following solution 
for $\eta^2_p$-components
\be
\eta^2_p=C\beta^{p-1}
\prod_{j=1}^{p-1}(1-\alpha^j)^{-\frac{1}{2}}\;,
\label{zeroeta}
\ee
where
\[
\beta=\frac{\lambda-\mu_f/2}{\lambda+\mu_f/2}\;.
\]
and $C$ is a normalization constant. This solution is well-behaved at
large $p$ for $\mu_f>0$. 
Note that zero mode \eq{eta1}, \eq{zeroeta} is
chiral
\[
\sigma_3\eta=-\eta\;.
\]
A zero mode with opposite chirality is localized on the antiwall
configuration. 

Let us work out the profile of zero mode along $z$-direction. This
profile is gauge invariant and can be regardedd as a probe
of the domain
wall shape. We assume $\mu, \mu_f\ll\lambda$. 
First, let us note that the  energy of zero mode on the $N$-th diagonal
is given by
\[
\omega=\frac{N-\gamma-1/2}{\rho}\;.
\]
Thus, the genuine Kaluza-Klein number of this mode is 
$N_0=N-\gamma-1$. From Eq.~\eq{zeroeta} one obtains the following
gauge invariant density
\be
\eta^{\dag}\eta=\sum_{p=1}^{\infty}\left(\eta^2_p\right)^2
\ket{N_0+\gamma+p}\bra{N_0+\gamma+p}\;.
\label{density}
\ee 
Its Weyl symbol is shown in Fig. 3.
\begin{figure}[tb]
\begin{center}
\epsfig{file=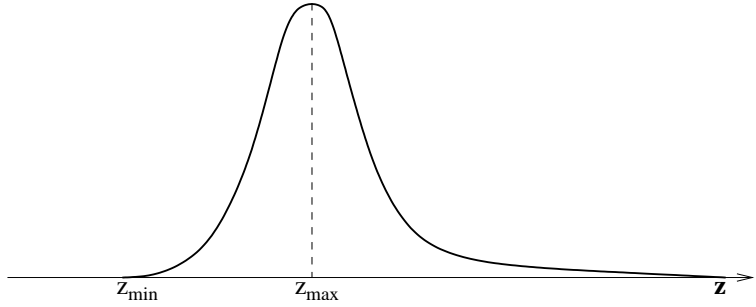,height=4.cm,width=10.cm} 
\end{center}
\caption{Gauge invariant 
profile of zero fermion mode localized on the domain wall.}
\end{figure}
This density concentrates near $p=p_{max}$ with $p_{max}$ determined by
\[
1-\alpha^{p_{max}}=\beta^2
\]  
or, explicitly,
\be
p_{max}=\frac{\lambda}{2\mu}\ln{\frac{\lambda}{2\mu_f}}
\label{pmax}
\ee
Coefficients entering Eq.~\eq{density} may be approximated 
in the vicinity of $p_{max}$ as follows,
\be
\left(\eta_p^2\right)^2=\tilde{C}^2\exp{\left(-\frac{2\mu_f\,\mu}{\lambda^2}
(p-p_{max})^2\right)}
\label{approxeta}
\ee
where $\tilde{C}$ is a constant. 
From 
Eqs.~\eq{density} and \eq{approxeta}, one 
infers the following properties of the Weyl symbol
$\widetilde{\eta^{\dag}\eta}(z)$ in terms of physical parameters: it
is equal to zero at $z<z_{min}$, where
\be
z_{min}=l\gamma+lN_0\;,
\label{zmin}
\ee
peaks in the vicinity of 
\be
z_{max}=\frac{1}{m}\ln{\frac{1}{2m_f\,l}}+l\gamma+lN_0\;.
\label{zmax}
\ee
where it has the form
\be
\widetilde{\eta^{\dag}\eta}(z)=\tilde{C}^2
\exp{\Bigl(-m_f\,m(z-z_{max})^2\Bigr)}\;,
\label{Weyleta}
\ee
and falls off exponentially at $z\to +\infty$.
The coordinates $z_{min}$ and $z_{max}$ can serve as two alternative
definitions of the position of the domain wall. 
If $N_0$ is not too large, the last terms in Eqs.~\eq{zmin},~\eq{zmax} can be
neglected. Then, $z_{min}\approx l\gamma$ 
coincides with our previous naive estimate
of the domain wall position, based on gauge dependent quantities (see
discussion after Eq.~\eq{chidisp}). 

On the other hand, $z_{max}$ is more appropriate for the definition of
the position of the wall, if one considers 
$\widetilde{\eta^{\dag}\eta}(z)$ as the gauge independent
shape of the domain wall seen by the zero fermionic mode. 
The disadvantage of this definition is that it 
depends on the mass of the fermion, but it is problematic to
provide a probe independent meaning to the notion of the 
domain wall shape.  
Assuming $m_f\approx m$, one obtains from Eqs.~\eq{zmax},~\eq{Weyleta}
the following
estimates for the position $z_{max}$ and width $\delta z$ of the domain wall
\be
z_{max}\approx \frac{1}{m}\ln{\frac{1}{2ml}}+l\gamma~,~~~
\delta z\approx \frac{1}{m}\;.
\label{position}
\ee
Let us stress once more that these estimates refer to the domain wall
as seen by the zero fermionic mode, and other probes may, in
principle, give other results. 
The estimate for the width agrees with that
deduced from the (gauge dependent) profile of the domain wall, see
Eq. \eq{dmwwidth}.


\section{Wall--antiwall system}
\label{wall-antiwall}

Similarly to the domain wall solution, it is convenient to describe
the wall--antiwall
configuration 
in terms of operators acting on a direct sum of Hilbert spaces
$H=H_1\oplus H_2$. 
Subspaces $H_1$ and $H_2$ emerging in the case of  the wall--antiwall
system are spanned by the
following systems of basis vectors
\begin{gather}
\ket{c_n}=\ket{n}~,~~~n\leq 0
\nonumber\\
\ket{h_p}=\ket{p}~,~~~p\geq 1\;.
\nonumber
\end{gather}
In these notations the wall--antiwall field has the following form
\bseq
\label{wallanti}
\begin{align}
&Z=\sum_{n=-\infty}^{0}(n+\gamma_1)\ket{c_n}\bra{c_n}
+\sum_{p=1}^{\infty}(p+\gamma_2)\ket{h_p}\bra{h_p}\label{wallantiZ}\\
&X_+=\sum_{n=-\infty}^{0}\sqrt{1-\alpha^{-n}}\ket{c_{n+1}}\bra{c_n}
+\sum_{p=1}^{\infty}\sqrt{1-\alpha^p}\ket{h_{p+1}}\bra{h_p}\label{wallantiX+}
\end{align}
\eseq
This may be viewed as the union of two fingerstalls. Two physically
different situations occur depending on the value of the parameter
$\Delta\gamma\equiv\gamma_2-\gamma_1$: for $\Delta\gamma > 0$ the two
fingerstalls intersect (see Fig. 4a) and, as we will see below, $U(2)$
gauge theory emerges in the region between the walls, while for
$\Delta\gamma < 0$ the fingerstalls are disconnected (Fig. 4b).

Due to the direct sum structure of Eqs.~\eq{wallanti}, they describe an
exact solution of field equations. This is somewhat unexpected, because it
implies that there is no attraction between the wall and antiwall. On
the other hand, configuration \eq{wallanti} belongs to the trivial
topological sector (it is straightforward to check that its topological
charge $Q$ defined in \eq {topcharge} is equal to zero), and, 
consequently, it is expected to be
unstable. We will show this explicitly later on.  

\begin{figure}[htb]
\begin{center}
\begin{picture}(500,100)(0,0)
\put(10,20){\epsfig{file=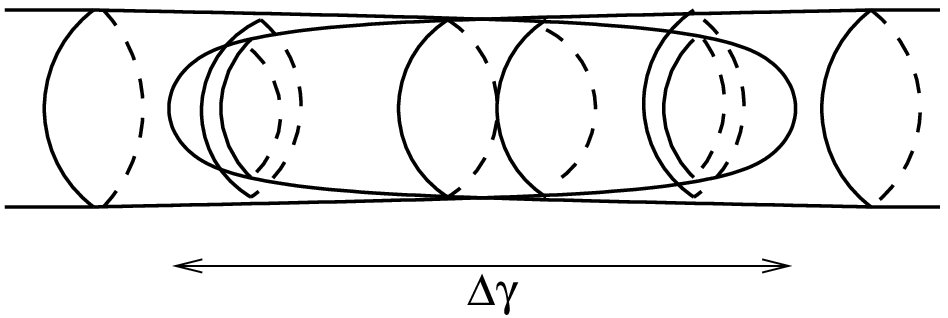,height=2.cm,width=5.cm} }
\put(230,20){\epsfig{file=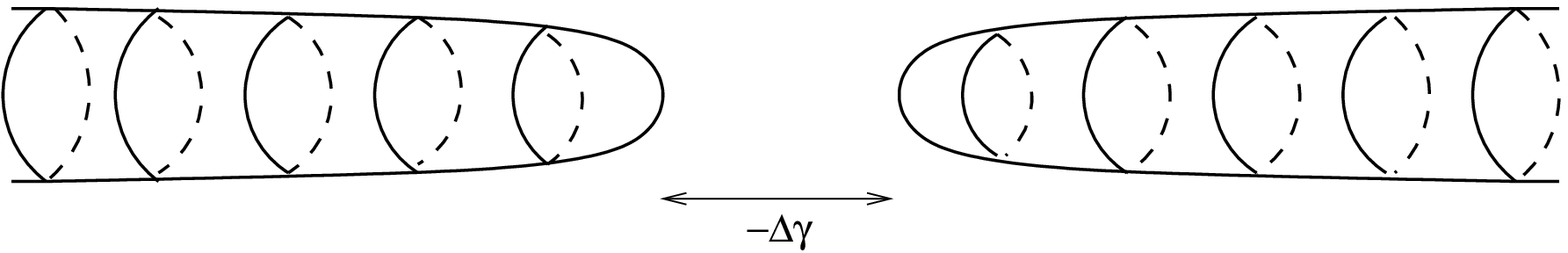,height=2.cm,width=8.cm}}
\put(82,0){a}
\put(341,0){b}
\end{picture}
\end{center}
\caption{Two possibilities for the non-BPS solution of two
fingerstalls:
a) Two regions with $U(1)$ theories separated by a region with $U(2)$
theory;
b) Two disjoint regions with $U(1)$ theories}
\end{figure}

To justify our interpretation of
the two-fingerstall configuration shown in Fig. 4a as a 
wall--antiwall system,
let us address a question of the 
restoration of non-Abelian symmetry in the region where 
the fingerstalls overlap.
When the size of this region is large, 
\be
\Delta\gamma\gg\frac{1}{|\ln\alpha|}\;,
\label{separation}
\ee
we expect $U(2)$ gauge symmetry to emerge 
in the overlap region. Direct analysis of
gauge field fluctuations in the background \eq{wallanti} is rather
cumbersome. More illuminating, as in the case of the single domain
wall, is to consider an adjoint scalar field $\phi$. It obeys 
Eq.~\eq{sceq} with the background given by Eqs.~\eq{wallanti}. For
simplicity we consider massless field, $m_\phi^2=0$. In analogy
to the single domain wall case, we
perform  the decomposition \eq{phidecomp}. Dispersion relations for modes
$\phi_{nm}$ and $\varphi_{pq}$ are again given by Eq.~\eq{disp}. These are
modes living on each of the fingerstalls separately. Modes
$\phi_{nm}$ do not feel the presence of the domain wall, but are
reflected back from the antiwall, while modes $\varphi_{pq}$ do not
feel the antiwall but are reflected by the wall. 
The same qualitative picture is valid for the modes of the gauge
fields responsible for $U(1)$ gauge symmetry in the asymptotic regions 
$z\to -\infty$ and $z\to +\infty$. This means that
the gauge fields from $z\to -\infty$
region  cannot penetrate into $z\to +\infty$ region, and vice versa. 

Modes $\phi_{nm}$
and $\varphi_{pq}$ correspond to diagonal elements of the $U(2)$ multiplet
in the region between the walls.
Off-diagonal components of the multiplet 
should come from the $\chi_{np}$ sector. Equation for these modes
reads
\begin{multline}
\rho^2\omega^2\chi_{np}=\left((p-n-\Delta\gamma)^2+2\lambda^2
\left(1-\frac{\alpha^p+\alpha^{p-1}+\alpha^{-n}+\alpha^{-n+1}}{4}
\right)\right)\chi_{np}\\-
\lambda^2\sqrt{(1-\alpha^{p-1})(1-\alpha^{-n+1})}\,\chi_{n-1,p-1}-
\lambda^2\sqrt{(1-\alpha^p)(1-\alpha^{-n})}\,\chi_{n+1,p+1}\;.
\nonumber
\end{multline} 
Performing the Kaluza-Klein decomposition, $p-n=N$,
$\chi_{p+N,p}\equiv\chi_p$ , one obtains
\begin{multline}
\rho^2\omega^2\chi_p=\left((N-\Delta\gamma)^2+2\lambda^2
\left(1-\frac{\alpha^p+\alpha^{p-1}+\alpha^{N-p}+\alpha^{N-p+1}}{4}\right)
\right)\chi_p\\-
\lambda^2\sqrt{(1-\alpha^{p-1})(1-\alpha^{N-p+1})}\,\chi_{p-1}-
\lambda^2\sqrt{(1-\alpha^p)(N-p)}\,\chi_{p+1}\;.
\label{chipeq}
\end{multline} 
The index $p$ in Eq.~\eq{chipeq} runs from $1$ to
$N$ because $n\leq 0$. If 
 the condition \eq{separation} is satisfied, low energy modes are
those with 
\[
|N_0|=|N-\Delta\gamma|
\ll\Delta\gamma
\] 
and their dispersion 
relation is approximately given by Eq.~\eq{disp} (with $N_0$ instead of
$N$). Strictly speaking, the wave vector $k$ of these modes is not
arbitrary and takes discrete values determined by the boundary conditions
at $p=1, N$. But the gap between two neighboring values is of order
\[
\Delta k\approx \frac{\pi}{l\Delta\gamma}
\]
and tends to zero as $\Delta\gamma\to\infty$. Thus, in the limit of
large separation between the walls, the spectra of the off-diagonal and
diagonal modes coincide, and $U(2)$ gauge symmetry emerges 
in the region between the walls. 

Now, let us return to the question of stability of the wall--antiwall
solution. As there is no force between the wall and antiwall,
instability, if any, must reveal itself in the existence of a tachyonic
mode in the spectrum of fluctuations of gauge fields about the
wall--antiwall 
background. Clearly, this tachyon should carry
indices corresponding to both fingerstalls comprising the solution \eq{wallanti}.
Thus, we consider fluctuations of the form
\[
Z=Z^{(0)}-a_{\theta}~,~~~X_+=X_+^{(0)}+b_+
\]
where $Z^{(0)}$, $X_+^{(0)}$ are given by Eqs.~\eq{wallanti}, and
$a_{\theta}$, $b_+$ are off-diagonal
\begin{align}
&a_{\theta}=\sum(a_{pn}\ket{h_p}\bra{c_n}+a^*_{pn}\ket{c_n}\bra{h_p})
\nonumber\\
&b_+=\sum(b_{pn}\ket{h_p}\bra{c_n}+c_{np}\ket{c_n}\bra{h_p})\;.\nonumber
\end{align}
Insertion of these expressions into  
Eq.~\eq{energyfunc} yields quadratic energy functional for the
fluctuations. The latter can be expressed as the sum of contributions
coming from different diagonals $p-n-1=N$. The contribution 
due to the 
$N$th diagonal is (we drop the overall dimensional prefactor
$\frac{2\pi}{g^2\rho l}$ in the energy)
\be
\begin{split}
{\cal E}_N=\sum_{p=1}^{p=N+1}\Biggl(&\left|(N-\Delta\gamma)b_p-
\sqrt{1-\alpha^{N-p+1}}\,a_p
+\sqrt{1-\alpha^{p-1}}\,a_{p-1}\right|^2\\
+&\left|(N-\Delta\gamma)c^*_p+\sqrt{1-\alpha^{p-1}}\,a_p
-\sqrt{1-\alpha^{N-p+1}}\,a_{p-1}\right|^2\\
+\frac{\lambda^2}{2}&\left|\sqrt{1-\alpha^{p-1}}\,c^*_p+
\sqrt{1-\alpha^{N-p+1}}\,b_p-\sqrt{1-\alpha^{N-p}}\,c^*_{p+1}
-\sqrt{1-\alpha^p}\,b_{p+1}\right|^2\\
+\frac{\mu^2}{2}&\left|\sqrt{1-\alpha^{p-1}}\,c^*_p+
\sqrt{1-\alpha^{N-p+1}}\,b_p+\sqrt{1-\alpha^{N-p}}\,c^*_{p+1}
+\sqrt{1-\alpha^p}\,b_{p+1}\right|^2\\ 
&~~~~~~~~~~~~~~~~~~~+\mu\lambda(\alpha^{p-1}+\alpha^{N-p+1})|c_p|^2
-\mu\lambda(\alpha^{p-1}+\alpha^{N-p+1})|b_p|^2\Biggr)\;.
\end{split}  
\label{energyN}
\ee
Notations in this expression are
\begin{align}
&b_p=b_{p,-N+p-1}&p&=1,2,~\ldots~,N+1\nonumber\\
&a_p=a_{p,-N+p}&p&=1,2,~\ldots~,N\nonumber\\
&c_p=c_{-N+p,p-1}&p&=2,3,~\ldots~,N\;.\nonumber
\end{align}
The expression \eq{energyN} is rather lengthy, 
and we presented  it just to demonstrate
that the only negative contribution to the energy comes from the modes
$b_p$. This suggests that these modes give the dominant contribution
to the tachyon. 

The following analysis is different for the two cases shown in
Fig. 4. Let us consider first the case $\Delta\gamma <0$. We argue
that in this case the lowest eigenvalue of the energy is given by the
contribution of the zeroth diagonal. Indeed, modes along other
diagonals have larger
Kaluza-Klein energy (which is proportional to $(N-\Delta\gamma)^2$ 
and $N\geq 0$). This effect shifts upwards masses squared of these modes.
Thus, let us study the case $N=0$. The expression \eq{energyN} is greatly
simplified yielding
\[
{\cal E}_0=\Bigl((\Delta\gamma)^2-2\mu\lambda\Bigr)|b_1|^2\;.
\]   
We see that $b_1$ is an eigenmode with the frequency 
\[
\omega^2=\frac{(\Delta\gamma)^2}{\rho^2}-\frac{m}{l}\;.
\]
This means that at
$\Delta\gamma<0$, the system is unstable, provided that
$|\Delta\gamma|<\rho\sqrt{m/l}$, but becomes stable at large 
values of $|\Delta\gamma|$. This behavior is fairly natural, because 
when $\Delta\gamma<0$ and $|\Delta\gamma|$ is large, 
two fingerstalls comprising the wall--antiwall 
system do not intersect and are well separated (Fig. 4b). 

Let us now turn to the case $\Delta\gamma > 0$.
If $\Delta\gamma$ satisfies Eq.~\eq{separation}, 
there is a large region
of intersection of the two fingerstalls with $U(2)$ 
gauge theory inside (Fig. 4a). 
One expects the system to be able to roll down to either $U(1)$ 
or $U(2)$ vacuum on the whole cylinder. Instead of trying to find the
tachyonic mode exactly, let us present an Ansatz which
demonstrates that the energy functional given by Eq.~\eq{energyN} 
has negative directions. To this end, let us choose $N=\Delta\gamma$, take
$a_p=0$, $c_p=0$ and rewrite the energy ${\cal E}_{N=\Delta\gamma}$ in the
following form
\be
\label{energygamma}
\begin{split}
{\cal E}_{N=\Delta\gamma}=\sum_{p=1}^{N/2}\frac{1}{2}
\left((\lambda+\mu)^2\left|\alpha\sqrt{1-\alpha^{N-p+1}}\,b_p
-\sqrt{1-\alpha^p}\,b_{p+1}\right|^2
-4\lambda\mu\alpha^{N-p+1}|b_p|^2\right)\\
+\sum_{p=N/2+1}^{N+1}\frac{1}{2}
\left((\lambda+\mu)^2\left|\sqrt{1-\alpha^{N-p+1}}\,b_p
-\alpha\sqrt{1-\alpha^p}\,b_{p+1}\right|^2
-4\lambda\mu\alpha^{p-1}|b_p|^2\right)\\
-2\lambda\mu(1-\alpha^{N/2})|b_{N/2+1}|^2\;.
\end{split}
\ee
Positive terms in eq.~\eq{energygamma} can be set to zero by the following
choice
\be
b_p=
\begin{cases}
\alpha^{p-1} \prod\limits_{j=1}^{p-1} 
\sqrt{\frac{1-\alpha^{N-j+1}}{1-\alpha^j}}&p=1,~\ldots~,N/2\\\\
\alpha^{N+1-p}\prod\limits_{j=1}^{p-1}
\sqrt{\frac{1-\alpha^{N-j+1}}{1-\alpha^j}}~~~~~~&p=N/2+1,~\ldots~,N+1\;.
\end{cases}
\label{bansatz}
\ee    
This Ansatz for $b_p$ is a symmetric combination of two
bell-shaped functions localized around $p_{max}$ and $(N-p_{max})$,
where $p_{max}$ is given by Eq.~\eq{pmax} (with $\mu$ instead of
$\mu_f$). Substitution of the Ansatz \eq{bansatz} into the
energy functional \eq{energygamma} yields
the following estimate for the tachyon energy
\[
\omega^2=-C\,\frac{4\mu^2}{\rho^2}\,\alpha^{\Delta\gamma-2p_{max}}=
-Cm^2\e^{-m\Delta z}\;,
\]
where $\Delta z=l(\Delta\gamma-2p_{max})$ is the separation between
the wall and antiwall, and $C$ is a coefficient of order one. We see
that the tachyon mass is exponentially small when the walls are far
away from
each other. 

This tachyon is not directly related to the distance between
domain walls $\Delta\gamma$ which is an exact modulus of the wall-antiwall
solution. One may guess that tachyon condensation leads to the change
of the shapes of the walls. The natural candidates for the end points of the
tachyon condensation are $U(1)$ and/or $U(2)$ vacua. We leave the study
of this condensation for future.

\section{M(atrix) theory interpretation of the domain walls}
\label{matrix}
The purpose of this section is to suggest a way of embedding the domain
wall solutions constructed in this paper into the matrix model of
M-theory
(see, e.g., Ref. \cite{Taylor} for a review of the M(atrix) model). Matrix
model is supersymmetric quantum mechanics described by the following
Lagrangian
\be
\label{1*}
L={1\over 2R}\Tr\left\{ \dot{\bf X}^i\dot{\bf X}^i+{1\over
2}[{\bf X}^i,{\bf X}^j]^2+(\mbox{fermions})\right\}\;,
\ee
where ${\bf X}^i$ ($i=1,\dots,9$) are real-valued $N\times N$ matrices
subject to the constraint
\[
[\dot{\bf X}^i,{\bf X}^i]=0\;.
\]
Originally \cite{deWit:1988ig}, this quantum mechanical system was
suggested as a regularized theory of a (super)membrane in 
flat 11-dimensional space-time in light-cone gauge. 
The matrices ${\bf X}^i$ play the role of embedding
functions of the membrane. Regularization is removed by taking the
limit $N\to \infty$. In this language $R=2\pi l_{11}^3$ is the membrane
tension.

Alternatively, one may consider Lagrangian (\ref{1*}) as an effective
low-energy description for a system of $N$ D0-branes in the type
IIA-theory in the $A_0=0$ gauge. In this case $R$ has an
interpretation of compactification radius of 11-dimensional M-theory
to ten dimensions and in string units, $l_s=1$, this radius
 is equal to the string
coupling $g_s$.

It was conjectured in Ref.~\cite{BFSS}, that large-$N$ limit of
the matrix model (\ref{1*}) describes all
of the M-theory in the infinite momentum frame. Moreover, it was
argued \cite{Sen:1997we,Seiberg:1997ad} that the 
quantum mechanical system (\ref{1*}) 
at finite $N$ describes a sector of M-theory with $N$ units of
momentum along compact light-like direction.

In Ref. \cite{Bak:2001kq} fuzzy cylinder was obtained as a BPS-solution in
the matrix model (in $A_0={\bf Z}$ gauge) and was interpreted as a D2-brane
of type IIA theory. In Ref. \cite{Bak:2001gm} field configurations
similar 
to our
Eqs. (\ref{conf}) were discussed in this context and were
interpreted as junctions of D2-branes. It is worth noting, however,
that these junctions (domain walls) were not obtained as solutions of
matrix model equations.

Here we would like to suggest that domain walls studied in this paper
may be obtained as solutions of the matrix model in {\it curved}
backgrounds. 

If one sets 
${\bf X}^i=0$ for $i\geq 4$
in the matrix model Lagrangian, one arrives at the action very
similar to the action (\ref{gaugeaction}) of gauge theory on the fuzzy
cylinder. The only difference is that extra terms linear in $\bf{X}$
and $\bf{Y}$ present in the definition of the field strength in this
theory (see Eqs. (\ref{F13}), (\ref{F23})) are absent in the 
matrix model. So our
purpose in this section is to find a way to introduce these terms, as well
as the term (\ref{addaction}), into the matrix model Lagrangian.

A generalization of the matrix model
Lagrangian (\ref{1*}) to arbitrary curved background is not known
(see Ref.~\cite{Douglas:1997ch} for a discussion of this problem). However, 
there is a proposal \cite{Taylor:1999gq} on how to modify the 
Lagrangian of the matrix
model to incorporate the effect of arbitrary  weakly curved 
background independent of the light
cone coordinate $x^-$. Namely,
to describe the effect of non-trivial eleven-dimensional metric
$g_{MN}=\eta_{MN}+h_{MN}$
and three-form $A_{MNL}$ at the linear level, one adds the
following terms to the Lagrangian of the matrix model
\begin{eqnarray}
\Delta L_g&=&{1\over 2}\sum{1\over
n!}\d_{i_1}\dots\d_{i_n}h_{MN}\;\mbox{STr}\l T^{MN}{\bf X}^{i_1}\dots
{\bf X}^{i_n}\r\label{Lg}\\
\Delta L_A&=&\sum{1\over
n!}\d_{i_1}\dots\d_{i_n}A_{MNL}\;\mbox{STr}\l J^{MNL}{\bf X}^{i_1}\dots
{\bf X}^{i_n}\r\label{LA}
\end{eqnarray}
where STr stands for the totally symmetrized trace
\[
\mbox{STr}\l A_{a_1}\dots A_{a_n}\r ={1\over
n!}~~~\Tr\!\!\!\!\!\!\!\!\!\!\sum_{\sm{transmutations}\;\sigma}
A_{\sigma(a_1)}\dots
A_{\sigma(a_n)}\;.
\]
We have not written terms describing magnetic interactions of the
membrane and terms with fermions. The former do not appear in the
backgrounds considered below, while the latter are not relevant for
our purposes.

Explicit expressions for the components of the energy-momentum tensor
$T^{MN}$ and antisymmetric current $J^{MNL}$ can be found in
Ref.~\cite{Taylor:1999gq}.
In what follows we will make use of the expressions for  the $T^{++}$
and $J^{+ij}$ components
\begin{eqnarray}
\label{T++}
T^{++}&=&{1\over R} \\
J^{+ij}&=&-{i\over 6R}[{\bf X}^i,{\bf X}^j]\;.\nonumber
\end{eqnarray}
Eqs. (\ref{T++}) and \eq{Lg} suggest that a natural starting point to construct
the M-theory background leading to extra terms like
(\ref{addaction}) in the matrix model action, is to consider metric
with non-trivial $g_{++}$ component. In order to be a legitimate
background of the M-theory at least in the supergravity approximation,
this metric should be supplemented with  the appropriate three-form
field to satisfy the equations of eleven-dimensional
supergravity. Thus one naturally arrives at the following class of the
supergravity solutions \cite{Hull:vh}
\begin{align}
&ds^2=-2dx^+dx^-+\sum dx^ldx^l-H(x^l)(dx^+)^2
\nonumber\\
&F_{+ijk}=\xi_{ijk}(x^l)\;,
\nonumber
\end{align}
where  functions $H(x^l)$ and $\xi_{ijk}(x^l)$ are related as follows
\be
\label{Heq}
\d_i^2H={1\over 6}\xi_{ijk}\;\xi^{ijk}
\ee
and 
\[
\xi\equiv {1\over 6}\xi_{ijk}dx^i\wedge dx^j\wedge dx^k 
\]
is closed and co-closed form.
These solutions are generalizations of the homogeneous pp-wave solutions
\cite{Berenstein:2002jq} which have attracted much attention recently.

To start with, let us present an example of the background admitting fuzzy
cylinder as a solution of field equations. For this purpose it
suffices to consider quadratic function $H(x^i)$. Say, one 
considers $H(x^i)$ of the following form
\bseq
\label{15}
\be
\label{15*}
H_{fc}(x^i)=l^2((x^1)^2+(x^2)^2)
\ee
supplemented with the following three-form
\be
\label{15**}
\xi_{fc}=l\,d\omega~, ~~~~~\omega=\l x^1dx^2\wedge dx^3-x^2dx^1\wedge dx^3\r\;.
\ee
\eseq
Applying the rules described above, it is straightforward to check that
the bosonic part of the matrix model Lagrangian in this background has
the following form
\be
\label{fuzzym}
L_{fc}={1\over 2R}\Tr\left( \dot{\bf X}^i\dot{\bf X}^i+
([{\bf X}^1,{\bf X}^3]+il{\bf X}^2)^2+([{\bf X}^2,{\bf X}^3]-il{\bf
X}^1)^2+[{\bf X}^1,{\bf X}^2]^2+\dots\right)\;,
\ee
where dots stand for non-negative terms in the potential, vanishing
for zero ${\bf X}^4,\dots,{\bf X}^9$. 
One immediately recognizes that the expression (\ref{fuzzym})
is the same as the Lagrangian of the gauge theory on the fuzzy 
cylinder where ${\bf X}^1,{\bf X}^2,{\bf X}^3$ play the role of 
covariant coordinates
${\bf X},{\bf Y},{\bf Z}$ and ${\bf X}^4,\dots,{\bf X}^9$ 
are massless adjoint scalar fields with a
specific positive definite quartic potential.

To construct gravitational background leading to the matrix model
admitting domain wall solutions discussed above, let us note
first that for two pp-wave solutions described by functions
$H_{1,2}$ and three-forms $\xi_{1,2}$, the sum $H_1+H_2$ and
$\xi_1+\xi_2$ is again a solution, if there are no hyperplanes with
non-vanishing fluxes for both three-forms $\xi_{1}$ and $\xi_{2}$. 
We have already described gravitational
background leading to the gauge theory on the fuzzy cylinder,
so now we have to find the function $H_m$ and three-form $\xi_{(m)}$ 
such that $\xi_{(m)123}=0$, which give rise to 
extra term of the form (\ref{addaction}) in the matrix model action.

A natural guess would be to take
\[
H_m={\mu^2\over\lambda^2}\l(x^1)^2+(x^2)^2-1\r^2\;.
\]
However, the Laplacian of this function is not positive definite,
\[
\d_i^2H_m=8{\mu^2\over\lambda^2}((x^1)^2+(x^2)^2-2)\;,
\]
in contradiction to Eq.~(\ref{Heq}). 

To get around this difficulty, one may
consider background depending on larger number of coordinates. For
instance, one may make the following choice of the function $H_m$ and
three-form $\xi_{(m)}$
\begin{align}
&H_m={\mu^2\over\lambda^2}\l(x^1)^2+(x^2)^2-(x^4)^2-(x^5)^2-1\r^2
\nonumber\\
&\xi_{(m)}=2\sqrt{2}\,{\mu\over\lambda}\,
d\omega~,~~~~~~\omega=(x^1x^4dx^2\wedge dx^5+x^2x^5dx^1\wedge dx^4)
\nonumber
\end{align}
The extra piece in the matrix model potential coming from this
background is 
\be
V_m={\mu^2\over 2\lambda^2}\mbox{STr}\l({\bf X}^1)^2+({\bf
X}^2)^2-1\r^2+\dots
\label{Vm}
\ee
where dots stand for the terms which are at least second order in
coordinates ${\bf X}^4,{\bf X}^5$. These terms do not affect equations 
for 
configurations with ${\bf X}^4={\bf X}^5=0$ which we are focusing on
here. Now, it
is straightforward to check that
\be
\label{18*}
{\mu^2\over 2\lambda^2}\mbox{STr}\l({\bf X}^1)^2+({\bf X}^2)^2-1\r^2=
{\mu^2\over
2\lambda^2}\Tr\l({\bf X}^1)^2+({\bf X}^2)^2-1\r^2+
{\mu^2\over 6\lambda^2}\Tr[{\bf X}^1,{\bf X}^2]^2+\dots
\ee
where dots now stand for commutator terms which do not affect the
field equations (but, in the $N\to\infty$ limit, in general contribute
to the energy). The first term in Eq.~(\ref{18*}) coincides with the
extra term given by Eq.~(\ref{addaction}) 
while the second one can be eliminated by the
redefinition of the parameters $\mu$ and $\lambda$ (at
$\mu^2<\lambda^2$).
Thus, Eq.~(\ref{18*}) provides an example of the supergravity
background leading to the matrix model with domain wall solutions
discussed in this paper. It is clear from the discussion above 
that one can construct a variety of backgrounds with this property.

There is a subtlety conserning the stability of the domain wall solution
in the matrix model. As we showed in section \ref{3}, if the extra
coordinates ${\bf X}^4,\dots,{\bf X}^9$ are disregarded,
the domain wall saturates the BPS bound \eq{BPSenergy}, and thus it is
stable. However, when the extra coordinates are included, the omitted
terms in Eq. \eq{Vm} may become tachyonic, in the domain wall
background, along the coordinates ${\bf X}^4,\dots,{\bf X}^9$, so that
the stability may be lost. In order to
ensure the stability, it is desirable to find a background 
admitting the domain wall solution, which saturates a BPS-type bound
for the {\em full} matrix model action. 

Such background can be obtained using technique proposed in
Ref.~\cite{Bonelli:2002fs}. The following class of pp-wave 
backgrounds with four
extra\footnote{Generally, pp-wave background leaves unbroken 16
supersymmetries out of 32 supersymmetries present in the
eleven-dimensional supergravity. These supersymmetries disappear after
gauge fixing of the kappa-symmetry in the supermembrane action, so the
resulting matrix model is not supersymmetric. In some cases pp-wave
may have extra unbroken supersymmetries, and the
corresponding matrix model is expected to be supersymmetric.} 
supersymmetries was described there,
\bseq
\label{B}
\begin{align}
&H=\left|{\d W(\phi)\over\d\phi^a}\right|^2
\label{BH}
\\
&\xi={1\over 4}d\l\epsilon_{abc}{\d
W(\phi)\over\d\phi^a}\d\phi^{+b}\wedge\d\phi^{+c}+h.c.\r
\label{Bxi}
\end{align}
\eseq
where $a=1,2,3$ and
\begin{align}
&\phi^1=x^1+ix^4\nonumber\\
&\phi^2=x^2+ix^5\nonumber\\
&\phi^3=x^6+ix^7\nonumber\;.
\end{align}
Also, it was suggested in Ref.~\cite{Bonelli:2002fs} that the matrix model
potential for this background has the following form\footnote{We present
terms in the potential which do not contain $x^3,x^8,x^9$
coordinates. The latter terms are the same as in flat space.}
\be
V_{\phi}=\Tr\l{1\over 8}[\Phi^a,\Phi^{+a}]^2+{1\over 2}\left|{1\over
2}\epsilon_{abc}[\Phi^b,\Phi^c]+
\d_a \tilde{W}(\Phi)\right|^2\r\;,
\label{VBonelli}
\ee
where $\Phi^a,~ a=1,2,3$ are matrices corresponding to the coordinates
$\phi^a$, and superpotenial $\tilde{W}(\Phi)$ is defined as follows,
\be
\label{tildeW}
\tilde{W}(\Phi)=\mbox{STr}\, W(\Phi)\;.
\ee
For quadratic superpotentials the ordering prescription following from
Eqs.~\eq{VBonelli}, \eq{tildeW} agrees with that
defined by Eqs.~(\ref{Lg}), (\ref{LA})\footnote{In particular, by choosing
$W=l^2\l(\phi^1)^2+(\phi^2)^2\r$ one may obtain supersymmetric
realization of the fuzzy cylinder in the matrix model (with
coordinates on the fuzzy cylinder ${\bf X}=\mbox{Re}\,{\Phi_1}$, 
${\bf Y}=\mbox{Re}\,{\Phi_2}$,
${\bf Z}=\mbox{Re}\,{\Phi_3}$ and all other fields set equal to zero),
which apparently differs from that of Ref.~\cite{Bak:2001kq}.}.
For generic superpotential, the two ordering prescriptions
are different. We have nothing to say about this discrepancy
here. If one adopts ordering defined by Eqs.~\eq{VBonelli}, \eq{tildeW},
background siutable for our purpose is a sum (in the sense 
explained above) of the pp-waves described by (\ref{15}) and the 
pp-wave given by (\ref{B}) with superpotential  
\[
W_m(\phi)={\mu\over\lambda}\phi^3\l(\phi^1)^2+(\phi^2)^2-1\r\;.
\]
It is straightforward to check that in this background the domain wall
\eq{domwall}
emerges as BPS solution (with identifications ${\bf X}=\mbox{Re}\,{\Phi_1}$,
${\bf Y}=\mbox{Re}\,{\Phi_2}$ and ${\bf Z}=x^3$ and 
all other coordinates set equal
to zero). It is worth noting, however, that $(2,2)$ supersymmetry is
explicitly broken in this background 
by terms coming from Eqs.~(\ref{15}).

\section{Summary and discussion}
\label{discussion}
Let us summarize our
results and discuss some open problems. 
In this paper we constructed and studied domain walls between
vacua with different  gauge groups $U({\cal N}_1)$ and $U({\cal N}_2)$
on one of the simplest NC
manifolds, fuzzy cylinder. We demonstrated that these domain walls
are characterized by a non-trivial topological charge and satisfy
BPS-like equations, provided an extra term stabilizing the radius
of the fuzzy cylinder is added to the action. 
They represent a novel class of exact NC gauge solitons in the sense that they
cannot be obtained by making use of solution generation technique 
of Refs. \cite{Polychronakos,Bak,Aganagic:2000mh,Harvey:2000jb}, and
do not have commutative counterparts. 

By making use of 
the  adjoint scalar and fundamental fermion fields as probes, we
studied some of the properties of the domain
walls and demonstrated that these objects exhibit rich
pattern of non-trivial phenomena. 
Namely, we addressed a question, whether fields 
charged under the gauge groups can penetrate from one side of the wall
to the other. 
The result is that if ${\cal N}_1<
{\cal N}_2$, fields charged
under $U({\cal N}_1)$ freely penetrate through the domain wall
into the $U({\cal N}_2)$ region where they become a part
of $U({\cal N}_2)$ multiplet. On the other hand, $U({\cal N}_2)$
fields which are not part of the $U({\cal N}_1)$ subalgebra experience
total reflection from the domain wall. It is worth mentioning here,
that the higher the mass (KK number) of the $U({\cal N}_2)$ mode, the
deeper it penetrates into the region with the $U({\cal N}_1)$ vacuum. 
This effect is an illustration of the UV/IR mixing characteristic 
to NC theories.

For fermion field we found that there is a zero mode localized on the
domain wall. Wave-function profile of this mode is a gauge invariant
characteristic of the shape of the domain wall. It would be
interesting to understand whether there is a NC analogue of the index theorem
relating the existence of this zero mode to non-trivial topological
properties of the domain wall.

Also we studied a non-BPS wall--antiwall configuration, dividing
the cylinder into three regions with $U(1)$, $U(2)$ and again
$U(1)$ gauge theories. A somewhat
unusual property of this system is the absence of the interaction
potential between wall and antiwall at the classical level. Still,
we found a tachyonic mode in the spectrum of perturbations about
this system, which we expect to roll down to either 
$U(1)$ or $U(2)$ vacuum on the whole cylinder. The
precise mechanism of the tachyon condensation deserves further study.
The mass of the tachyon becomes exponentially small at large
separation between the wall and antiwall.

There is a simple brane picture for the domain walls between
$U({\cal N}_1)$ and $U({\cal N}_2)$ theories. Namely, one can think
of this configuration as a stack of ${\cal N}_2-{\cal N}_1$ 
branes of fingerstall shape inserted into ${\cal N}_1$
cylindrical branes. Some of the properties discussed above have a
natural interpretation in the D-brane language. 
For instance, the pattern of  
gauge field spectrum in the presence of the domain wall
is quite transparent in the string context. 
Indeed, gauge fields belonging to the $U({\cal N}_1)$ multiplet
correspond to strings with both ends on the cylindrical branes; these
strings do not feel the presence of the domain wall at all. Strings
corresponding to other gauge fields, on the contrary, have at least one
end on the fingerstall branes and are bound to them, so that they
cannot propagate along the entire cylinder.  
On the other hand, we are
not aware of the stringy interpretation of the localized fermion
mode. 

We suggested a way to embed the $(U({\cal N}_1)-U({\cal N}_2))$ domain
walls into
M-theory. Namely, we suggested that they emerge as solutions of the
matrix theory corresponding to the curved supergravity backgrounds
which have
the form of pp-waves with inhomogeneous three-form field strength. 
This effect can be thought of as
a generalization of the Myers effect \cite{Myers:1999ps}. 
We considered two ways of introducing effects of curved backgrounds
into the matrix model: one based on calculations at weak curvature,
and the other relying on supersymmetry. These approaches lead to
apparently different prescriptions in our case, however they both yield
backgrounds admitting the domain wall solutions. The advantage of the
latter approach is that it leads to the domain wall solution which
saturates the BPS-type bound for the {\em full} matrix model action,
and thus is stable. 
It is worth noting that our arguments rely on the approximation of weakly
curved background; one may hope that they may apply beyond this
approximation, especially taking into account that pp-wave
backgrounds similar to those we discussed here were
shown to be exact string backgrounds
\cite{Russo:2002qj,Bonelli:2002fs}.
We leave aside an issue of the supersymmetrization of the domain walls,
though the BPS property suggests that it should be possible.

To conclude, 
results obtained in this paper demonstrate that the rank of the gauge group
can be a non-trivial dynamical parameter in the NC gauge theories.
\section*{Acknowlegements}
We are indebted to V. Rubakov for active collaboration at the
initial stage of this work and for subsequent stimulating interest and
fruitful discussions. We are grateful to D. Bak for drawing our
attention to gauge theories on fuzzy cylinder. We thank G. Bonelli and
A. Konechny for useful correspondence. We appreciate fruitful
discussions with F. Bezrukov, D. Gorbunov, D. Levkov and G. Rubtsov.    
S.S. would like to thank for hospitality the DESY Theory Group, where
a part of this work was done. 
This work has been supported in part by Russian Foundation for Basic
Research, grants 02-02-17398 and Swiss Science Foundation grant
7SUPJ062239.  The work of S.D. has
been supported in part by INTAS grant YS 2001-2/128. The work of S.S. has
been supported in part by INTAS grant YS 2001-2/141.

\end{document}